\documentclass[12pt]{article}
\usepackage[utf8]{inputenc}
\usepackage[english]{babel}
\usepackage{amssymb,graphicx}
\pdfoutput=1
\usepackage{graphics}
\usepackage{amsmath,amsfonts}
\usepackage{fullpage}
\usepackage{wrapfig} 
\usepackage{braket}
\usepackage[colorinlistoftodos]{todonotes}
\usepackage{multirow}

\title{\vspace{-2cm} 
\begin{flushright}
{\normalsize INR-TH-2020-044}
\end{flushright}
\vspace{0.5cm} 
Examining axion-like particles with superconducting radio-frequency cavity}
\author{Dmitry Salnikov$^1$\thanks{{\bf e-mail}: salnikov.dv16@physics.msu.ru}, Petr Satunin$^2$\thanks{{\bf e-mail}: satunin@ms2.inr.ac.ru}, D.~V.~Kirpichnikov$^2$\thanks{{\bf e-mail}: kirpich@ms2.inr.ac.ru}, Maxim Fitkevich$^{2,3}$\thanks{{\bf e-mail}: fitkevich@phystech.edu} \vspace{.2cm}\\
\normalsize\it $^1$ Moscow State University, \\ 
\normalsize \it Leninskiye Gory, 119991 Moscow, Russia\\
\normalsize\it $^2$ Institute for Nuclear Research of the Russian Academy
of Sciences, \\  
\normalsize \it  60th October Anniversary Prospect, 7a, 117312  Moscow, Russia\\
\normalsize\it $^3$ Moscow Institute of Physics and Technology, \\  
\normalsize \it  Institutskiy per., 9, 141701 Dolgoprudny, Moscow Region, Russia}
\date{}
\begin{document}
\maketitle

\begin{abstract}
We address production of massive axion-like particles by two electromagnetic modes inside a superconducting radio-frequency (SRF) cylindrical cavity. We discuss in detail the 
choice of pump modes and cavity design. We numerically compute time-averaged energy density of produced axion field 
for various cavity modes and wide range of axion masses. This allows
us to estimate optimal conditions for axion production within a 
cavity. In addition, we consider photon regeneration process 
initiated by produced axion field in a screened radio-frequency 
cavity and derive 
constraints in parameter space $(g_{a\gamma\gamma},\ m_a)$ for 
different choice of pump modes. 
\end{abstract}

\section{Introduction}


Axion-like particles (ALPs) are hypothetical pseudoscalar particles appearing in 
several extensions of the Standard Model \cite{Irastorza:2018dyq}. Original axions 
were introduced in order to resolve the strong CP-problem in QCD 
\cite{Peccei:1977hh,Peccei:1977ur}. Later, it was argued that  ALPs can appear in a
low-energy phenomenological description of string theory 
\cite{Svrcek:2006yi,Arvanitaki:2009fg}. 

The efforts toward the searches for ALPs include such types of experiments as:
helioscopes~\cite{Anastassopoulos:2017ftl}, 
haloscopes~\cite{Kahn:2016aff,Asztalos:2001tf,JacksonKimball:2017elr}, 
light-shining-through-wall (LSW) 
experiments~\cite{Spector:2019ooq,Tan:2018tcs}, space-based gamma-ray telescopes~\cite{Egorov:2020cmx}, 
accelerator-based experiments~\cite{Feng:2018pew,Berlin:2018bsc,Dusaev:2020gxi,Banerjee:2020fue}, 
neutrino experiments~\cite{Brdar:2020dpr} and reactor experiments~\cite{Dent:2019ueq}.   

In addition, astrophysics and cosmology observations imply that ALPs are well 
motivated candidates for dark matter content \cite{Preskill:1982cy, Abbott:1982af, Dine:1982ah}. 
Moreover, several exotic scenarios of DM can be associated with 
ALPs~\cite{Berezhiani:1989fp, Berezhiani:1992rk}. 
A properties of axion dark matter are sensitive to the self-interaction parameters. In particular, the relevant dark matter
can be 
clumped into miniclusters \cite{Hogan:1988mp, Kolb:1993zz}, or form other inhomogeneous 
structures \cite{Vilenkin:1984ib,Sakharov:1994id, Sakharov:1996xg}. Axion field can also 
form exotic compact objects (bose stars \cite{Colpi:1986ye, Tkachev:1991ka}) providing a 
possible explanation for fast radio bursts \cite{Levkov:2020txo, Buckley:2020fmh}.

More generally, the axion field $a$ with mass $m_a$ and dimensionful coupling to 
photons $g_{a\gamma\gamma}$ is described by the Lagrangian
\begin{equation}\label{Lagrangian}
\mathcal{L}=-\frac{1}{4} F_{\mu\nu}F^{\mu\nu}+\frac12(\partial_\mu a)^2 - \frac12{m_a}^2 a^2 +\frac{g_{a\gamma\gamma}}{4}\,a\,F_{\mu\nu}\tilde{F}^{\mu\nu}\;,
\end{equation}
where $F_{\mu\nu}$ is the electromagnetic tensor and $\tilde{F}^{\mu\nu}=\epsilon^{\mu\nu\rho\sigma}F_{\rho\sigma}/2$ is its dual. 
The Lagrangian~\eqref{Lagrangian} yields the following 
equations for axion and electromagnetic fields,
\begin{equation}
(\partial_\mu \partial^\mu +{m_a}^2)\,a=\frac{g_{a\gamma\gamma}}{4} F_{\mu\nu}\tilde{F}^{\mu\nu}\, ,
\label{EqForAx}
\end{equation}
\begin{equation}
\label{Max-eqns}
\partial_\mu F^{\mu\nu}=g_{a\gamma\gamma}\,\tilde{F}^{\mu\nu}\partial_\mu a.
\end{equation}
If the electromagnetic invariant $F_{\mu\nu}\tilde{F}^{\mu\nu} = - 4 (\Vec{E}\cdot\Vec{B})$ is non-vanishing then Eq.~\eqref{EqForAx} implies that axion field 
can be produced.
This 
may be realized in laboratory by combination of two strong electromagnetic (EM) waves\footnote{For a monochromatic EM wave in vacuum $(\Vec{E}\cdot\Vec{B})$ vanishes.} 
or by a single EM wave in 
a strong magnetic field. Strong enough EM field with high level of coherence can be produced within optical range by lasers or within radio-frequency range inside SRF cavities.

The axion field once being produced may interact with the
EM field in a non-linear way according to Eq.~\eqref{Max-eqns}. 
So that the axion-induced EM field may be detected within the same production cavity~\cite{Bogorad:2019pbu} or within an additional detection cavity~\cite{Hoogeveen:1992nq,Janish:2019dpr,Gao:2020anb}. 
In the latter case both cavities should be screened in order to suppress the external EM field penetration. This setup illustrates so-called 
LSW type of laboratory experiments for axion searches. For instance, both production and detection cavities are filled with the strong magnetic field in order to initiate effective axion-photon conversion.

This setup was proposed and realized for both optical \cite{VanBibber:1987rq} and RF ranges \cite{Hoogeveen:1992nq}. 
Both optical LSW experiment ALPS \cite{Ehret:2010mh} and RF experiment
CROWS \cite{Betz:2013dza} gives the same order of magnitude constraints\footnote{The same order of magnitude constraint came
from optical polarization experiment PVLAS \cite{Zavattini:2012zs}.} (for details see  Ref.~\cite{Irastorza:2018dyq}) in the plane 
$(g_{a\gamma\gamma},m_a)$. In particularly, for small axion masses one has $g_{a\gamma\gamma} \lesssim 10^{-7}\,\mbox{GeV}^{-1}$, 
which is still the best pure laboratory constraint. Although  significantly better constraints (up to 
$g_{a\gamma\gamma} \lesssim 10^{-10}\, \mbox{GeV}^{-1}$) come from null results of dark matter searches 
or solar axion detection (see, e.~g.~Ref.~ \cite{Anastassopoulos:2017ftl}), these constraints are sensitive to the model of axion production. On the other hand, the production of ALPs in laboratory experiments is 
straightforward. However, both cosmic and laboratory methods for ALPs searches complement each 
others.  

The classical LSW setup requires external magnetic field in both  
production and detection cavities. However, the quality factor of cavities is constrained at level $Q\lesssim 10^5$ that implies the limitation on the amplitude of cavity modes. Therefore, sensitivity of ALPs detection decreases. The much bigger quality factor $Q\sim 10^{12}$ can be achieved with superconducting radiofrequency (SRF) cavities, but the price to be payed is that one can not apply strong magnetic field inside the cavity due to degradation of surepconducting state.
The maximal amplitudes for SRF cavity modes are constrained by the overall magnetic field near the cavity walls. In particular, for superconducting niobium \cite{Kittel} the critical magnetic field is $\sim 0.2$~T. Given that 
constraint, the authors of Ref.~\cite{Janish:2019dpr} suggested the LSW setup involving cylindrical SRF production cavity and toroidal SRF detector. Moreover, it was pointed out that sensitivity depends essentially on the geometrical formfactor for emitter and converter cavities.

In our paper we address this issue in detail. In particular, we discuss spatial distributions of the produced axion field in cylindrical SRF cavities. We also estimate detection sensitivity of the cylindrical RF cavity ($Q\sim 10^{5}$) filled with the strong magnetic field, so that it can reach $10$~T. Recently, a similar setup was suggested in Ref.~\cite{Gao:2020anb}, particularly, authors discuss the LSW facility to probe ALPs with two screened cylindrical SRF cavities which are served as emitter and receiver of axion field respectively. The setup allows to achieve relatively large quality factors $Q\sim 10^{12}$, however, the peak EM field in the cavities is constrained by $\lesssim 0.2$~T. We show that sensitivity to probe ALPs in our LSW setup $g_{a\gamma\gamma} \lesssim \mathcal{O}(1)\times 10^{-11}\, \mbox{GeV}^{-1}$ is comparable to that performed in Ref.~\cite{Gao:2020anb}.

In Ref.~\cite{Bogorad:2019pbu} authors suggested the setup to probe ALPs inside a single SRF cavity filled with different modes. Axion-like particles realize a non-linear coupling between those modes. 
Absence of the magnetic field allows to achieve the quality factor $Q\sim 10^{12}$. 
However, sensitivity is leveraged by relatively small magnitude of 
magnetic field $\lesssim 0.2$~T allowed in SRF. 
We show that our LSW setup is also sensitive to probe ALPs for the region of parameter space, which is close to one discussed in Ref.~\cite{Bogorad:2019pbu}.

This paper is organized as follows. In Sec.~\ref{sec:axion-production} we consider ALPs production using Green function approach. In Sec.~\ref{sec:num-results} we present 
results of numerical calculation for time-averaged 
energy density of axion field initiated by different pairs of cylindrical cavity eigenmodes. In Sec.~\ref{sec:detection} we consider ALPs detection in the RF cavity with the strong magnetic field. We also estimate constraints in the parameter space $(g_{a\gamma\gamma}, m_a)$.
In Sec.~\ref{sec:discuss} we discuss obtained results. Appendices contain technical details. 

\section{ALP production in superconducting cavity}\label{sec:axion-production}


In this section we consider production of the axion field  by electromagnetic radio-frequency modes pumped into a superconducting cavity. The generated axion
field is described by a solution of  Eq.~\eqref{EqForAx} respecting causality. In particular, it is given by
\begin{equation}\label{Green}
a(\Vec{x},t) = \int\limits_{-\infty}^{\infty} dt'
\int\limits_{V_{\mathrm{cav}}}d^3x'\,G_{\mathrm{ret}}(\vec{x}-\vec{x}',t-t')\;\times\; \left[  -g_{a\gamma\gamma}\,\left( \vec{E}(\vec{x}',t')\cdot \vec{B}(\vec{x}',t') \right)\right]\;,
\end{equation}
where $G_{\mathrm{ret}}(\vec{x}-\vec{x}',t-t')$ is the retarded Green function, 
 $\vec{E}(\vec{x}',t')$ and  $\vec{B}(\vec{x}',t')$ are the electric and magnetic 
 fields respectively inside the cavity of volume $V_{\mathrm{cav}}$. Since for a 
 single cavity mode the electric field is orthogonal to the magnetic one, at least 
 two cavity modes are necessary for ALPs productions. Therefore, one reads $\vec{E}(\Vec{x},t)=\vec{E}_1(\Vec{x},t)+\vec{E}_2(\Vec{x},t)$ and  $\vec{B}(\Vec{x},t)=\vec{B}_1(\Vec{x},t)+\vec{B}_2(\Vec{x},t)$ for the electric  and magnetic fields correspondingly,  where the subscripts refer to the cavity modes at given frequencies $\omega_{1,\,2}$. 

The time dependence for each mode decouples as follows 
$$ \vec{E}_i(\Vec{x},t)=\sqrt{2} \Re \mathrm{e} \left[ \Vec{{\cal E}}_i(\vec{x},\omega_i) e^{-i\omega_i t} \right]\;, \qquad\qquad \vec{B}_i(\Vec{x},t)=\sqrt{2} \Re \mathrm{e} \left[ \Vec{{\cal B}}_i(\vec{x},\omega_i) e^{-i\omega_i t} \right]\;, $$
where $\Vec{{\cal E}}_i(\vec{x},\omega_i)$
are cavity eigenmodes without proper normalization to take into account its arbitrary amplitudes. The electromagnetic invariant $(\vec{E}\cdot\vec{B})$ for two modes can be represented as
\begin{equation}\label{Fpm}
\left( \vec{E}(\vec{x},t)\cdot \vec{B}(\vec{x},t) \right) =  \Re \mathrm{e}\left[ F_+(\Vec{x})\cdot\mathrm{e}^{-i\omega_+ t} + F_-(\Vec{x})\cdot\mathrm{e}^{-i\omega_- t}\right]\;,
\end{equation}
where we defined $\omega_\pm=\omega_2 \pm \omega_1$ and
$$ F_+(\Vec{x}) \equiv \vec{{\cal E}}_1(\Vec{x}) \cdot \vec{{\cal B}}_2(\Vec{x}) + \vec{{\cal E}}_2(\Vec{x}) \cdot \vec{{\cal B}}_1(\Vec{x})\;, \qquad F_-(\Vec{x}) \equiv \vec{{\cal E}}_1^*(\Vec{x}) \cdot \vec{{\cal B}}_2(\Vec{x}) + \vec{{\cal E}}_2(\Vec{x}) \cdot \vec{{\cal B}}_1^*(\Vec{x})\;.
$$
Note that we consider cavity modes where the electric and magnetic fields are orthogonal, so that
$(\vec{{\cal E}}_i(\vec{x})\cdot\vec{{\cal B}}_i(\vec{x}))=0$ everywhere inside the cavity.

Since Eq.~(\ref{EqForAx}) is linear with respect to $a$, one has independent 
propagation for each frequency component in Eq.~(\ref{Fpm}). The produced axion 
field is then given by $a(\Vec{x},t)=a_+(\Vec{x},t)+a_-(\Vec{x},t)$, where 
$a_\pm(\Vec{x},t)$ are independent components for $\omega_\pm$. One can easily 
integrate out $t'$ in Eq.~\eqref{Green} for two cases depending on axion mass (see e.~g.~Appendix~\ref{app}),
\begin{equation}\label{GretTileSup2}
 m_a< \omega_\pm:\qquad \qquad a_\pm(\Vec{x},t) = -g_{a\gamma\gamma} \; \Re \mathrm{e}\,\int\limits_{V_{\mathrm{cav}}} d^3x'\,\frac{F_\pm(\Vec{x}')}{4\pi |\Vec{x}-\Vec{x}'|} \mathrm{e}^{-i\omega_\pm t+i|\Vec{x}-\Vec{x}'|k_\pm }\;.
\end{equation}
\begin{equation}\label{GretTileSup3}
 m_a> \omega_\pm:\qquad \qquad a_\pm(\Vec{x},t) = -g_{a\gamma\gamma} \; \Re \mathrm{e}\,\int\limits_{V_{\mathrm{cav}}} d^3x'\,\frac{F_\pm(\Vec{x}')}{4\pi |\Vec{x}-\Vec{x}'|} \mathrm{e}^{-i\omega_\pm t-|\Vec{x}-\Vec{x}'|\kappa_\pm }\;,
\end{equation}
where $k_\pm \equiv \sqrt{\omega_\pm^2-m_a^2}$. For the case $m_a>\omega_\pm$ Eq.~(\ref{GretTileSup3}) is obtained formally by a replacement of $i k_\pm$ in Eq.~(\ref{GretTileSup2}) by $\kappa_\pm = \sqrt{m_a^2-\omega_\pm^2}$, so that the axion field amplitude decreases exponentially far from the cavity volume.  
We note that the functions $F_\pm(\Vec{x}')$ are generally complex what may give an additional phase of the integrand in Eqs.~(\ref{GretTileSup2})-(\ref{GretTileSup3}).

Since the axion field amplitude harmonically oscillates the time-averaged $a_\pm(\vec{x},t)$ goes to zero. Instead, we consider the following time-averaging,
\begin{equation}
\braket{a^2}= \frac{1}{T}\int\limits^{T}_0 dt \,\, a^2(t) = \frac{1}{T} \int\limits^{T}_0 dt \,\, (a_+(t) + a_-(t))^2 = \braket{a_+^2} +\braket{a_-^2}\;, \label{eq:averaging}
\end{equation}
where the mixed term $\braket{a_+a_-}$ vanishes because $a_+a_-$ is a sum of products of two harmonic functions with different frequencies and averaging over time period yields zero.

For two non-zero terms in Eq.~(\ref{eq:averaging}) we obtain
\begin{align} 
 m_a< \omega_\pm:\qquad \qquad
      &\braket{a^2_\pm} = \frac{1}{2}\left(\left(A_\pm^C\right)^2 + \left(A_\pm^S\right)^2\right)\;,\\
 m_a>\omega_\pm:\qquad \qquad
      &\braket{a^2_\pm} = \frac{1}{2}B_\pm^2\;,
\end{align}
where
\begin{equation}
\label{ACAS}
A^{C(S)}_\pm = g_{a\gamma\gamma} \, \int\limits_{V_{\mathrm{cav}}} d^3x'\, 
    \frac{\left|F_\pm(\vec{x}')\right|}{4\pi|\Vec{x}-\Vec{x}'|}    \left\{\begin{array}{c}
\cos \left(k_\pm |\Vec{x}-\Vec{x}'|\right) \\
\sin \left(k_\pm |\Vec{x}-\Vec{x}'|\right)
\end{array} \right\}\;,
\end{equation}
\begin{equation}
\label{Bpm}
B_\pm = g_{a\gamma\gamma} \, \int\limits_{V_{\mathrm{cav}}} d^3x'\, 
    \frac{\left|F_\pm(\vec{x}')\right|}{4\pi |\Vec{x}-\Vec{x}'|} \mathrm{e}^{-\kappa_\pm |\Vec{x}-\Vec{x}'|}\;.
\end{equation}
For a specific resonant case $m_a = \omega_\pm$ one has $k_\pm=\kappa_\pm=0$, so that
\begin{equation}
\label{Ares}
\left. A^C_\pm \right|_{\mathrm{res}}= \left. B_\pm \right|_{\mathrm{res}} = g_{a\gamma\gamma} \, \int\limits_{V_{\mathrm{cav}}} d^3x'\, 
    \frac{\left|F_\pm(\vec{x}')\right|}{4\pi |\Vec{x}-\Vec{x}'|}\;, \qquad \left. A^S_\pm \right|_{\mathrm{res}}=0\;.
\end{equation}
Next, far away from the cavity the integral  $\int d^3x' \left|F_\pm(\vec{x}')\right| /|\Vec{x}-\Vec{x}'|$ is suppressed
 for two transversal magnetic modes in the cavity (TM+TM), 
 since  their overlap factor $\int d^3x' \left|F_\pm(\vec{x}')\right|$
 is negligible.  In that case the resonant production of ALPs in SRF cavity is
 ineffective. In addition we note that for two transversal electric modes 
 (TE+TE) or transversal magnetic and electric modes (TM+TE) this overlap integral can be significant. Therefore the resonant production of 
 ALPs for these combinations of modes is more efficient.
To conclude this section let us list the formulae 
for the energy density of the generated axion field,
\begin{equation}
\rho^E_\pm = \frac{1}{2}\dot{a}^2 + \frac{1}{2}(\partial_i a)^2  + \frac{m_a^2}{2}a^2\;. \label{eq:energy-den}
\end{equation}
 In particular, for various masses $m_a$ of the axion field the time-averaged value $\langle\rho^E_\pm\rangle$ can be written as follows,
\begin{align}
m_a < \omega_\pm:& \qquad
\braket{\rho^E_\pm} =\frac{1}{4}
\left(
\left[m_a^2+\omega^2_\pm\right]((A_{\pm}^S)^2+(A_{\pm}^C)^2)+  
(\partial_i A_{\pm}^S)^2+(\partial_i A_{\pm}^C)^2
\right)\;,   
\\
m_a> \omega_\pm:& \qquad
\braket{\rho^E_\pm} =\frac{1}{4}
\left(
\left[m_a^2+\omega^2_\pm\right]B_{\pm}^2+  
(\partial_i B_{\pm})^2
\right)\;,
\end{align}
where $A_{\pm}^{C(S)}$ and $B_\pm$ are given by 
Eqs.~(\ref{ACAS})-(\ref{Bpm}). In Sec.~\ref{sec:num-results} we discuss 
spatial distributions for the given quantities and study its properties for the 
resonant case $m_a \simeq \omega_+$.



\section{Numerical results for ALP production in cylindrical cavity}\label{sec:num-results}

In this Section we consider axion production in cylindrical cavity. We use 
TE$npq$/TM$npq$ notation to classify EM cavity modes \cite{Hill}. Given a height
$L$ and a radius $R$ of the cavity one has the following  dispersion relations 
for TM and TE modes respectively,
\begin{equation}
\omega^{TM}_{npq} =\sqrt{\left(\frac{x_{np}}{R}\right)^2 + \left( \frac{q\pi}{L} \right)^2}, \qquad\qquad \omega^{TE}_{npq} =\sqrt{\left(\frac{x'_{np}}{R}\right)^2 + \left( \frac{q\pi}{L} \right)^2}\;,
\end{equation}
where $x_{np}$ and $x'_{np}$ are p-th roots of the n-th order Bessel function $J_n(x)$ and its derivative $J'_n(x)$ correspondingly. Integers $n,\,p,\,q$ enumerate the full set of modes and refer to the ``winding''
numbers in $\phi,\,\rho,\, z$ directions respectively. The explicit expressions 
for the given set of modes are presented in Appendix~\ref{AppB}.


One has to consider the following combinations of pump cavity modes:
(i) TM+TM, (ii) TE+TE, (iii) TE+TM. 
It is straightforward to show using expressions from Appendix~\ref{AppB} that
for (i) and (ii) cases the functions $F_+$ and $F_-$ are purely imaginary and for 
the case (iii) $F_+$ and $F_-$ are purely real. Since ${\cal E}^{TE}_z={\cal B}^{TM}_z=0$ one has
\begin{equation}
\label{FTETM}
(TE+TM): \qquad \left| F_\pm\right| = \left| {\cal E}_z^{TM}{\cal B}_z^{TE} + {\cal E}_\rho^{TM}{\cal B}_\rho^{TE} + {\cal E}_\phi^{TM}{\cal B}_\phi^{TE} \pm \left( {\cal E}_\rho^{TE}{\cal B}_\rho^{TM}   + {\cal E}_\phi^{TE}{\cal B}_\phi^{TM} \right) \right|\;,   
\end{equation}
\begin{equation}
\label{FTT}
(TE+TE\text{ or }TM+TM): \qquad\left| F_\pm\right| = \left| {\cal E}_\rho^1{\cal B}_\rho^2 + {\cal E}_\phi^1{\cal B}_\phi^2 \pm \left( {\cal E}_\rho^2{\cal B}_\rho^1 + {\cal E}_\phi^2{\cal B}_\phi^1 \right) \right|\;.    
\end{equation}
Functions $F_\pm$ vanish for the cases (i) and (ii) as soon as its ``winding'' numbers $n_1=n_2=0$. On the other hand, two modes with zero $n$ give
non-vanishing terms for the case (iii).

We performed numerical calculations\footnote{Multidimensional numerical integration in \cite{Github} is based on the package \cite{package}.} \cite{Github}
of the time-averaged axion energy density $\braket{\rho^E_\pm}$ generated by two TE/TM modes in the cylindrical cavity with various dimensions $R$ and $L$. We also assumed $g_{a\gamma\gamma}=10^{-10}\,\mbox{GeV}^{-1}$ as a benchmark for ALPs coupling. One also has to fix the amplitudes ${\cal E}_0$ (${\cal B}_0$), which appears as a normalization constants in the expressions for mode components ${\cal E}_z^{TM}$ (${\cal B}_z^{TE}$) of Appendix~\ref{AppB}. Its maximum value is limited by a requirement that the magnetic field on the superconducting cavity walls should not exceed the critical value $\sim 0.1$ T. We note that the components ${\cal E}^{TM}_\rho$ (${\cal B}^{TE}_\rho$) and ${\cal E}^{TM}_\phi$ (${\cal B}^{TE}_\phi$) can be larger than ${\cal E}_0$ (${\cal B}$) for a ``pancake-like'' design of the cylindrical cavity with $R\gg L$. Therefore, in numerical calculations we require typical values $|\Vec{{\cal E}}|$, $|\Vec{{\cal B}}| \lesssim 0.1$~T for both TM and TE modes.


\begin{figure}[h!]
\begin{minipage}[h]{0.49\linewidth}
\center{\includegraphics[width=1\linewidth]{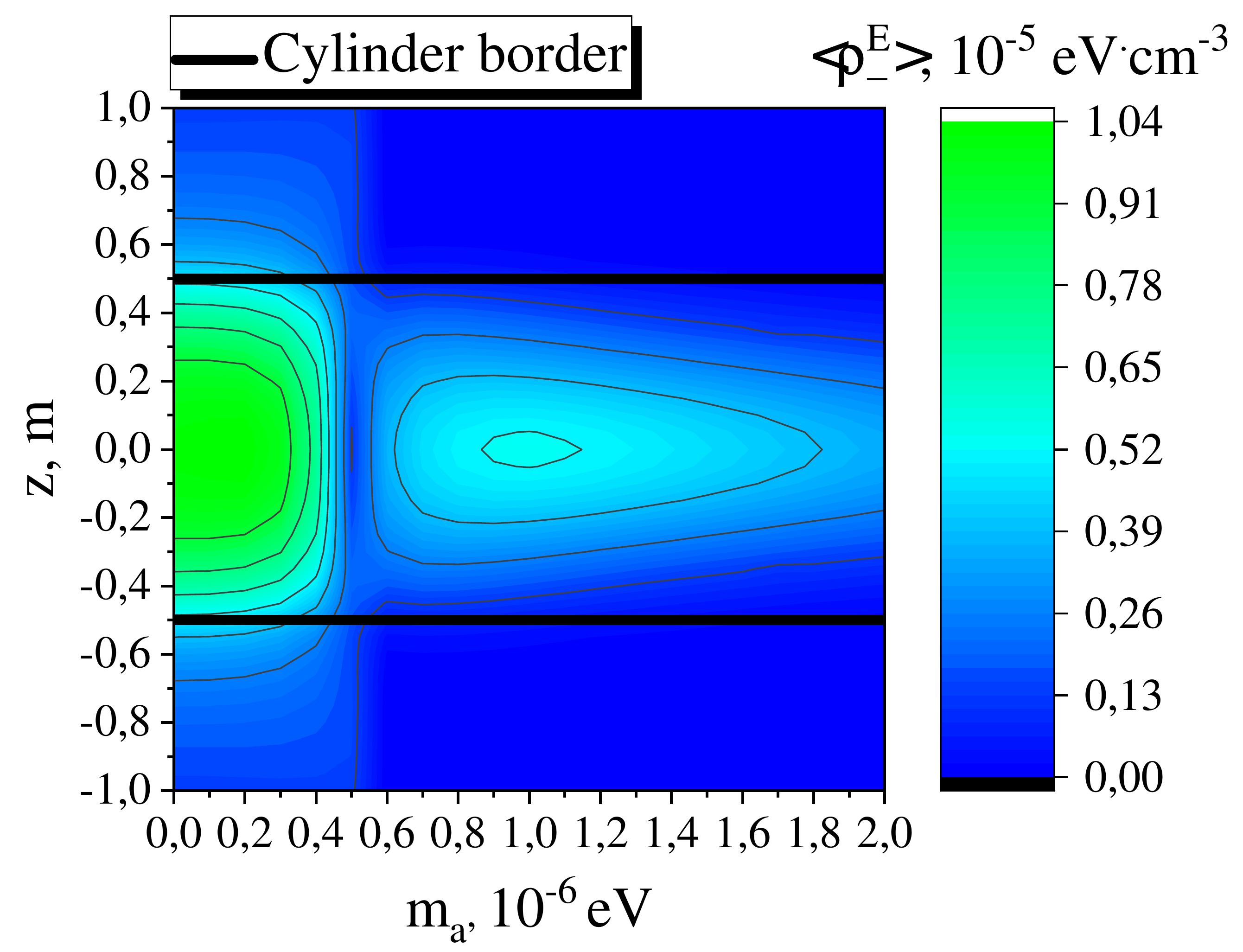}}
\center{\includegraphics[width=1\linewidth]{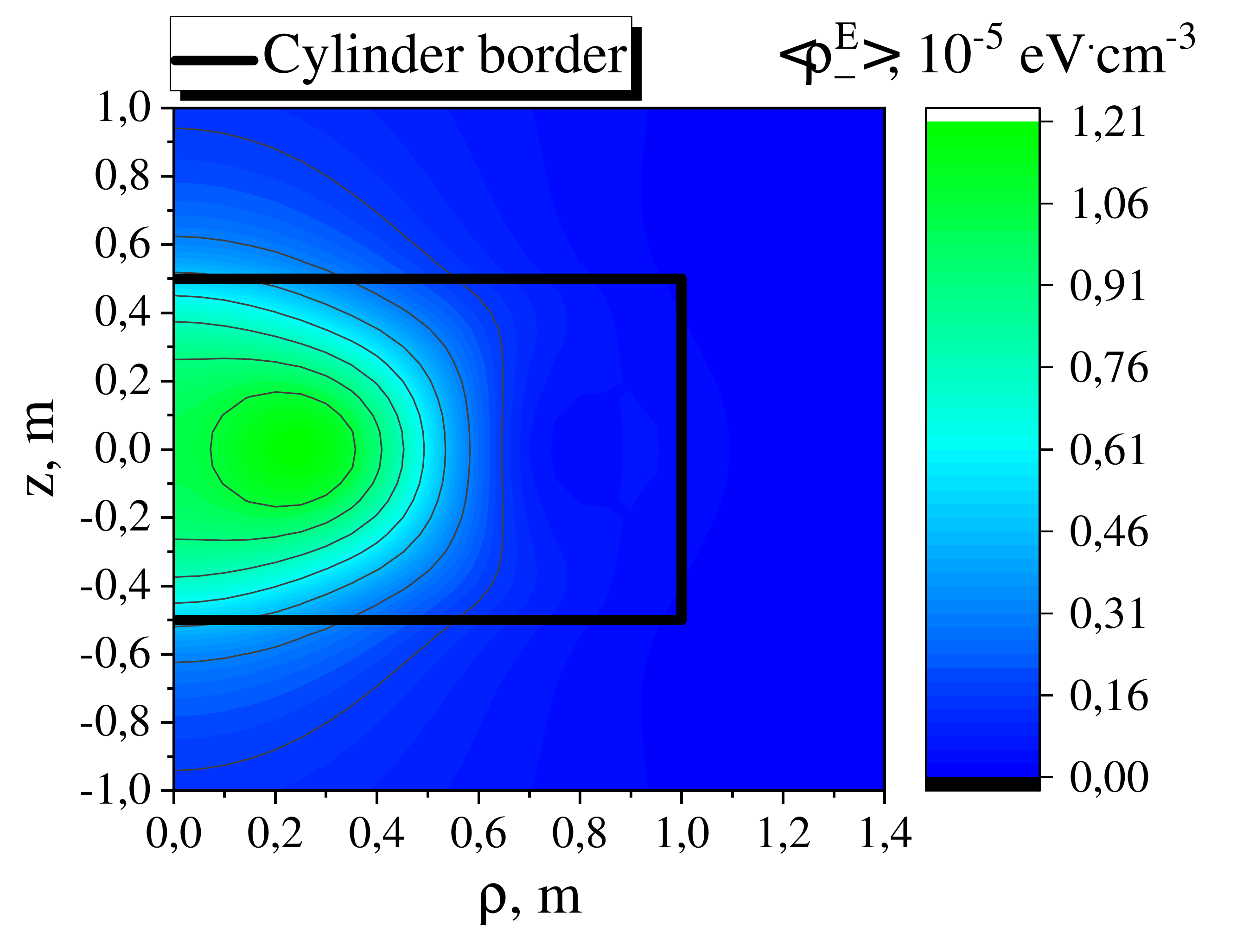}}
\end{minipage}
\hfill
\begin{minipage}[h]{0.49\linewidth}
\center{\includegraphics[width=1\linewidth]{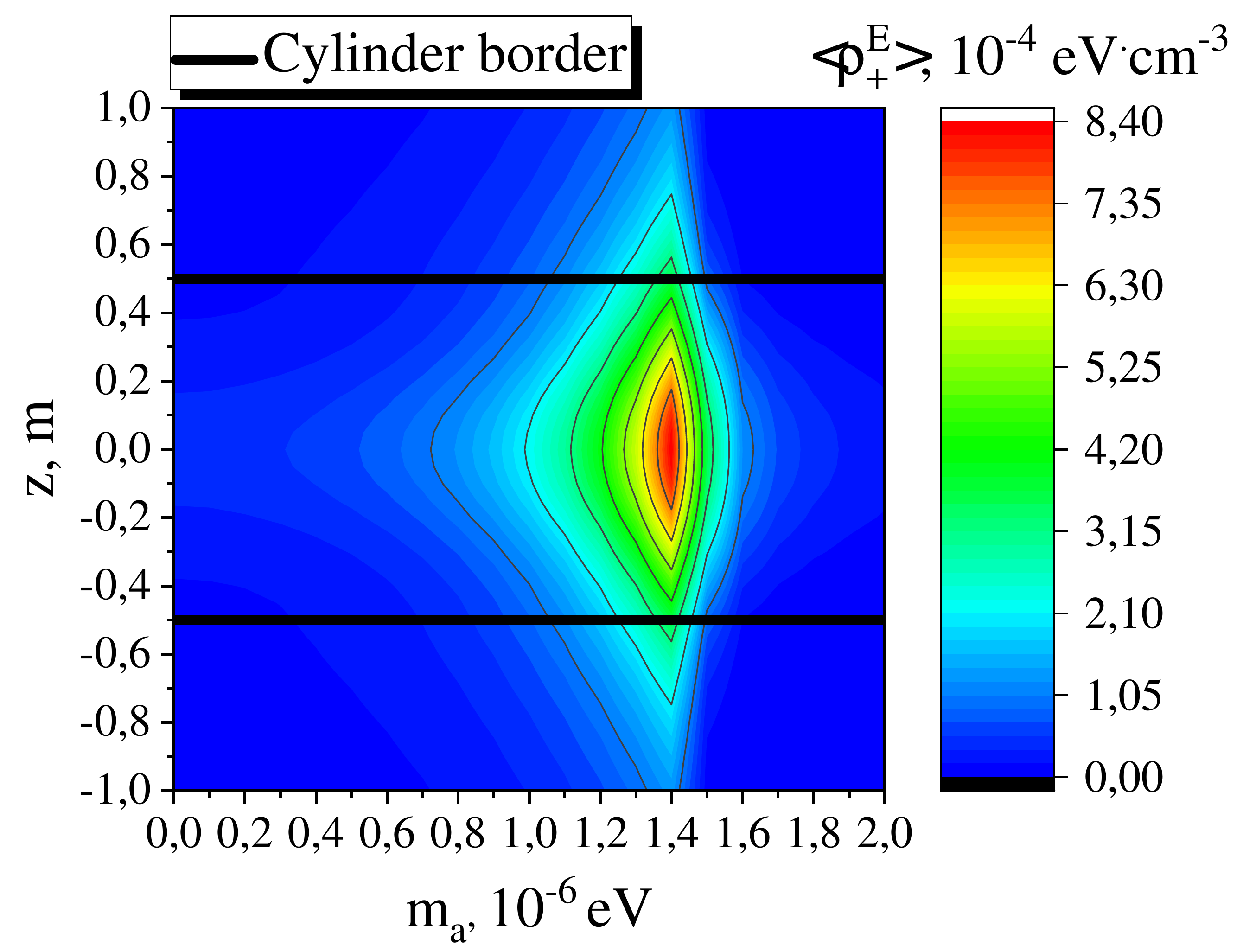}}\\
\center{\includegraphics[width=1\linewidth]{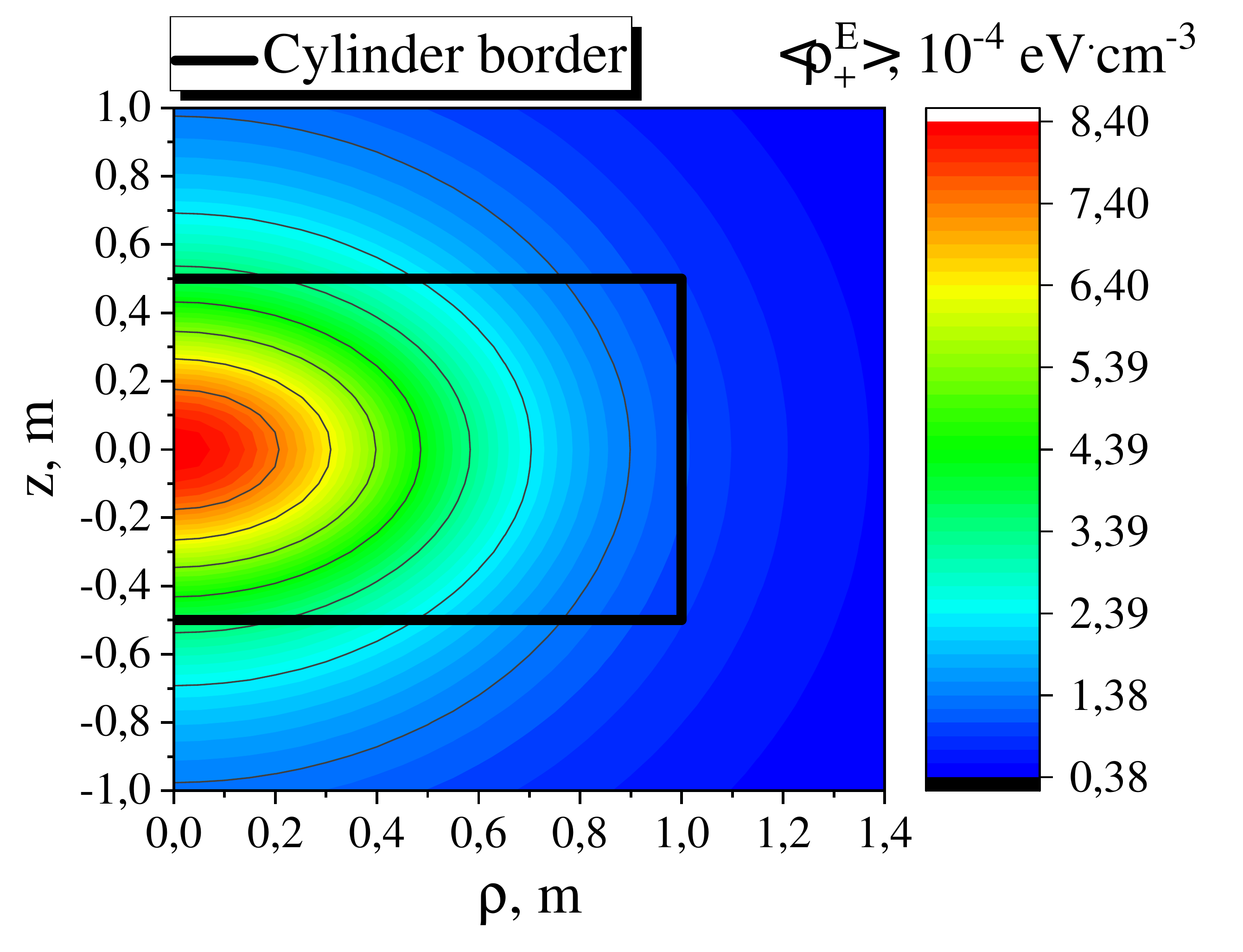}}
\end{minipage}
\caption{ Results of numerical calculations for TM010+TE011 pump modes. Top: Contour plots for the time-averaged energy densities $\braket{\rho^E_{-}}$ and $\braket{\rho^E_{+}}$ evaluated on the cylinder axis as function of axion mass $m_a$ and distance $z$ from center of cavity with TM010+TE011 pump modes. Cavity dimensions: $L = 1$ m, $R = 1$ m. Left panel: $\omega_-=5.1 \cdot 10^{-7}$ eV. Right panel: $\omega_+=14.7\cdot 10^{-7}$ eV. Bottom: Spatial distribution of the time-averaged energy density $\braket{\rho^E_-}$ and $\braket{\rho^E_+}$ on the cavity section along its axis $(\rho,\,z)$ with TM010+TE011 pump modes. Cavity dimensions: $L = 1$ m, $R = 1$ m. Left panel: $m_a=0$,  $\omega_-=5.1 \cdot 10^{-7}$ eV. Right panel: $m_a=\omega_+=14.7\cdot 10^{-7}$  eV.} \label{fig:1}
\end{figure}


Let us consider the production cavity with dimensions $R=L=1$~m and
the simplest combination of pump modes TM010+TE011. The evaluated axion energy density for each frequency component $a_\pm$ is presented in Fig.~\ref{fig:1}. Plots at the top in Fig.~\ref{fig:1} show the
time-averaged  energy density on the cylinder axis ($\rho=0$) as function of both distance from the cavity center $z$ and axion mass $m_a$. Top-right panel in 
Fig.~\ref{fig:1} shows the resonance in $\langle \rho^E_+ (m_a\simeq\omega_+)  \rangle$ in the center of cavity. On the other hand, on the top-left panel in Fig.~\ref{fig:1} one can see a significant suppression of $\langle \rho^E_-  \rangle$ in the center of cavity and relatively larger amplitude at  $m_a=0$.
That suppression can be explained as follows. One can see that $F_-$ in Eq.~(\ref{FTETM}) is a difference between two positive terms of the same order of magnitude, which compensate each other. 
However $F_+$ is a sum of the relevant terms. Therefore the energy density associated with $\omega_+$ is almost two orders of magnitude larger than that for $\omega_-$.

The spatial distribution of energy density as function of radius $\rho$ 
and height $z$ is shown at the bottom in Fig.~\ref{fig:1}. These plots  
respect an axial symmetry  since the chosen cavity modes do not depend on $\phi$. Distributions for $\omega_+$ and $\omega_-$ 
components of axion field were calculated for the cases
$m_a=\omega_+$ and $m_a=0$ respectively. For both 
cases the axion energy density is localized close to center of the production 
cavity and decreases outside the cavity. 
In particular, the energy density $\langle \rho_+^E \rangle$ near the cavity ends at $z=0.5$~m drops by factor 2 with respect to a resonant value at $z=0$~m.

The results of numerical calculations for other modes are shown in Figs.~\ref{fig:C1}-\ref{fig:C2}. For these modes we consider the ALPs density
$\langle \rho_+^E \rangle$ only, because $\langle \rho_-^E \rangle$ term is negligible. 
One has the resonances at $m_a \simeq \omega_+$ as expected. However, the intensity of the 
resonance may 
drastically depend on the combination of modes. In particular,
for certain combinations of 
modes the  axion production rate is suppressed by factor $\sim 10^{2}$ in 
comparison with other combinations.  This suppression occurs if (i) $q_1 + q_2$ is 
even, or (ii)  $n_1 \neq n_2$. In fact, in those two cases an overlap factor for two modes is zero,  $\int d^3x' F_\pm (x') = 0$ (see Appendix in 
Ref.~\cite{Berlin:2019ahk}). Therefore, for given modes the resonant amplitude 
(\ref{Ares}) tends to zero more rapidly far away from cavity.

\section{Detection}
\label{sec:detection}
\begin{figure}[t]
\center{\includegraphics[width=0.5\linewidth]{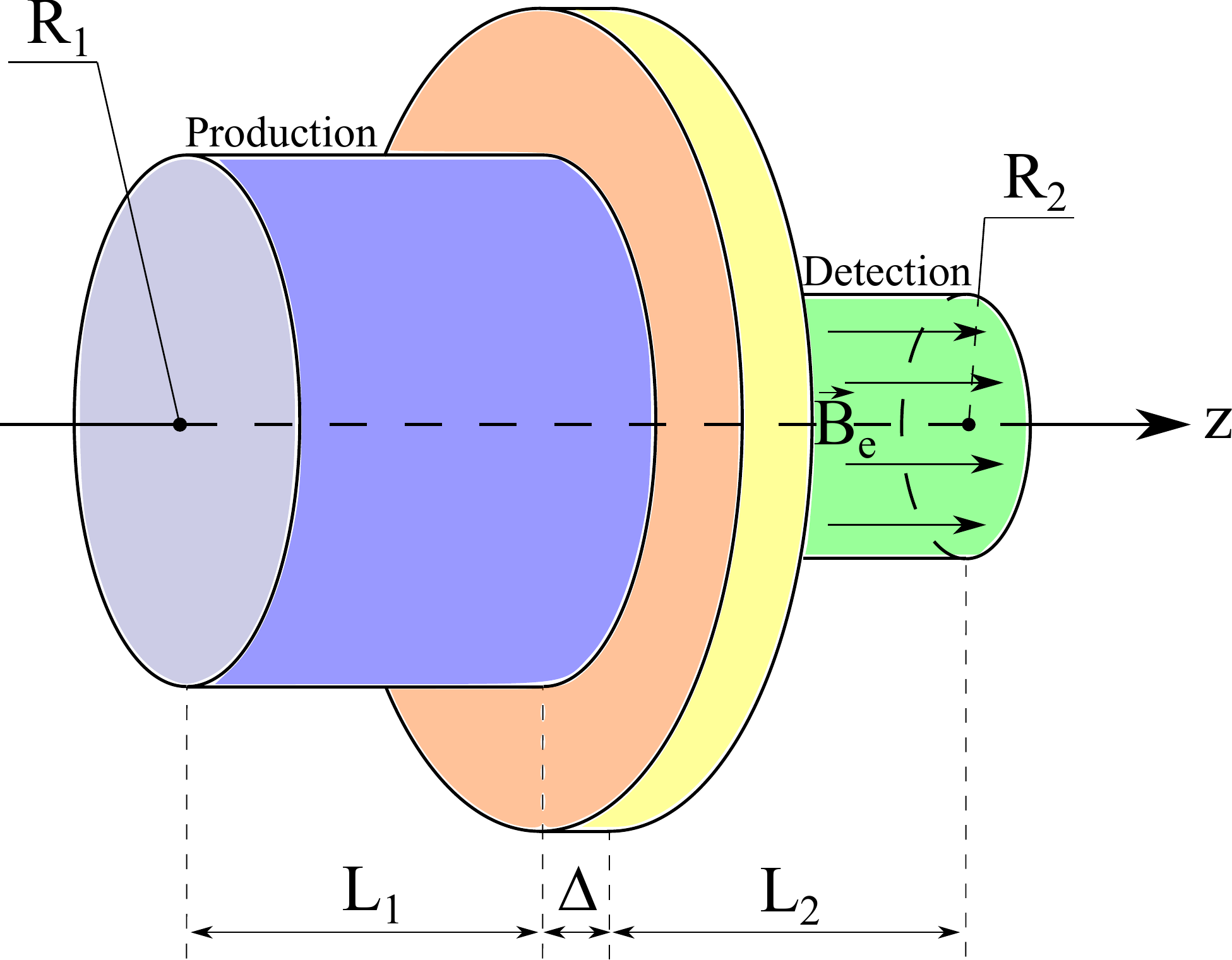}}
\caption{ A scheme for the setup considered. The detection cavity ($R_2$) is separated from the production cavity ($R_1$) by a screening plate of width $\Delta$. }
\label{fig:pic}
\end{figure}

In order to obtain some information about produced axions we have to include a second cavity as a detector in the setup\footnote{It is feasible to detect axions within the same cavity as it was proposed in Ref.~\cite{Bogorad:2019pbu}. However, we leave resonant detection of the ALPs for the  given setup to future.}. There are two detection options, in first one assumes that the detection cavity filled with the strong magnetic field (the setup similar to haloscope \cite{Sikivie:2020zpn}). The second option is associated with the oscillating electromagnetic field inside the detection cavity (see, e.~g. recent Ref.~\cite{Berlin:2019ahk}). The detection cavity in our setup is filled with the constant magnetic field $\Vec{B}_e(\Vec{x})$. In particular, we consider the setup with two coaxial cylindrical cavities separated by a screening plate of width $\Delta$, see Fig.\ref{fig:pic}, where indices $1$ and $2$ refer to the production and detection regions respectively.

The detection cavity response to the external axion field is determined by Eq.~\eqref{Max-eqns}. 
The latter one decouples into a pair of Maxwell's equations with axion-induced current,
\begin{equation}
    (\vec{\nabla}\cdot\Vec{E})=-g_{a\gamma\gamma}\,(\Vec{B}\cdot\vec{\nabla}\,a)\;, \qquad\qquad
    [\vec{\nabla}\times\Vec{B}]=\frac{\partial}{\partial t}\Vec{E}+g_{a\gamma\gamma}\left(\Vec{B}\frac{\partial}{\partial t}a-[\Vec{E}\times\vec{\nabla}\,a]\right)\;.
\end{equation}
The electric field of a given signal mode $\vec{E}^{\mathrm{sig}}(\Vec{x},t)$ inside the detection cavity obeys 
the following equation
\begin{align}
    \left( \frac{\partial^2}{\partial t^2} + \Gamma \frac{\partial}{\partial t} - \Delta \right)\Vec{E}^{\mathrm{sig}}(\Vec{x},t)=g_{a\gamma\gamma}\left(\vec{\nabla}\,(\vec{B}_e(\Vec{x})\cdot\vec{\nabla}\,a(\Vec{x},t))-\vec{B}_e(\Vec{x})\frac{\partial^2}{\partial t^2}a(\Vec{x},t)\right)
,
    \label{MainEqOnE}  
\end{align}
where we introduced the damping coefficient $\Gamma$ to take into account dissipation effects; r.h.s 
is associated with the ALPs and magnetic field $\Vec{B}_e(\vec{x})$ inside the cavity. 

Next, to simplify our considerations we take uniform magnetic field $\vec{B}_e$ directed along $z$ axis inside the cavity. 
We use a mode expansion for the signal electric field inside the detection cavity,
\begin{equation}
    \vec{E}^{\mathrm{sig}} (\Vec{x},t)=\sum_m\Vec{{\cal E}}_m(\Vec{x})E^{\mathrm{sig}}_m(t)\;,
    \label{SeriesEm}
\end{equation}
where $\Vec{{\cal E}}_m(\Vec{x})$ are cavity eigenmodes with fixed normalization,
\begin{equation}
   \Delta\,\Vec{{\cal E}}_m(\Vec{x})+\omega_m^2\Vec{{\cal E}}_m(\Vec{x})=0\;, \qquad \qquad \int\limits_{\mathrm{2\,cav}}d^3x\,(\Vec{{\cal E}}_m(\Vec{x})\cdot\Vec{{\cal E}}_n(\Vec{x}))=V_2\cdot \delta_{mn}\;,
   \label{norm}
\end{equation}
satisfying necessary boundary conditions. We note that index $m$ accounts for $n,\,p,\,q$ winding numbers of the 
cavity.  The integration in (\ref{norm}) is performed
over the volume  of the detection cavity $V_2$.
Since the signal modes are orthogonal, one obtains the following equation by
substituting \eqref{SeriesEm} into (\ref{MainEqOnE}) and multiplying each cavity mode by $\Vec{{\cal E}}_m(\Vec{x})$,

\begin{equation}
\label{DetectEq}
    \left(\frac{\partial^2}{\partial t^2}+\Gamma\frac{\partial}{\partial t}+\omega_m^2\right)E_m^{\mathrm{sig}}(t)=-g_{a\gamma\gamma}\int\limits_{\mathrm{2\,cav}}\frac{d^3x}{V_2}\left[(\Vec{{\cal E}}_m(\Vec{x})\cdot\Vec{B}_e)\,\ddot{a}(\vec{x},t) -  \left(\Vec{{\cal E}}_m(\Vec{x})\cdot \left(\vec{B}_e \vec{\nabla}\right)\vec{\nabla} a(\Vec{x},t)  \right)\right]\;,
\end{equation}
where the r.h.s is an axion-induced driven force. The second term in the r.h.s. of 
(\ref{DetectEq}) is suppressed compared to the first one if the produced axions are non-relativistic. 
However, in the general case these two terms seem to be of the 
same order. Since the magnetic field in the detection cavity $\vec{B}_e$ is directed along $z$ axis, the scalar product 
$(\Vec{{\cal E}}_m(\Vec{x})\cdot\Vec{B}_e)$ does not vanish if only ${\cal E}^z_m$ does not depend on $z$, or the winding number for the detection mode $q=0$. Moreover, it is more convenient to use the detection mode TM010. For that mode the 
only nonzero component is ${\cal E}^z_m$. The corresponding electric field, including normalization factor, reads ${\cal E}^z_{TM010}(\Vec{x}) = 1.92 \, J_{0}\left(\dfrac{x_{01}}{a}\rho\right) $.
Thus, (\ref{DetectEq})  simplifies,
\begin{equation}
\label{DetectEq1}
      \left(\frac{\partial^2}{\partial t^2}+\Gamma\frac{\partial}{\partial t}+\omega_m^2\right)E_m^{\mathrm{sig}}(t)= -g_{a\gamma\gamma}B_e^z\int\limits_{\mathrm{2\,cav}}\frac{d^3x}{V_2}{\cal E}_m^z(\vec{x})\,\left( \ddot{a}(\vec{x},t) - \partial_z^2 a(\vec{x},t)  \right)\,,
\end{equation}
If the r.h.s. of Eq.~(\ref{DetectEq1}) oscillates as $e^{-i\omega t}$, the complex solution of Eq.~(\ref{DetectEq1}) 
reads the r.h.s. multiplied to $ (-\omega^2 - i\omega \Gamma + \omega_m^2)^{-1}$. 
This is exactly the case of aforementioned produced axion field  determined by Eqs.~(\ref{GretTileSup2}) and~(\ref{GretTileSup3}). For 
$m_a < \omega_\pm$ the signal electric field  $E_m^{\mathrm{sig}}(t)$ reads
\begin{equation}
\label{DetectSol}
   E_m^{\mathrm{sig}}(t)=\frac{-g_{a\gamma\gamma}^2 B_e^z}{\omega_\pm^2-\omega_m^2 + i\omega_\pm \Gamma}   \int\limits_{\mathrm{2\,cav}}\frac{d^3x}{V_2}{\cal E}_m^z(\vec{x})\,\int\limits_{\mathrm{1\,cav}} d^3x'\frac{F_\pm(x')}{4\pi}\left( \omega_\pm^2+\partial_z^2\right) \frac{\mathrm{e}^{-i\omega_\pm t + i k_\pm|x-x'|}}{|x-x'|}\;.
\end{equation}
The signal field $E_m^{\mathrm{sig}}(t)$ has a narrow peak at $\omega_m \simeq \omega_\pm$. 
The frequency $\omega_m$ of cavity detection eigenmode  can be adjusted to a 
value $\omega_m \simeq \omega_\pm$ by fixing the radius $R_2$ of the 
detection cavity.
The damping coefficient $\Gamma$ can be expressed via the quality factor $Q$ of the
detection cavity, $\Gamma =  \omega_m/Q$. Thus, averaged squared amplitude for
electric field of the signal mode reads
\begin{equation}
  \langle |E_m^\mathrm{sig}(t)|^2\rangle 
  \equiv 1/2 \left(E^{\pm c}_m\right)^2+1/2 \left(E^{\pm s}_m\right)^2\;,  
\end{equation}
where
\begin{equation}
E^{\mathrm{\pm c(s)}}_m=\frac{g_{a\gamma\gamma}^2E_0^2Q {B}^z_c}{4\pi } \cdot 
\frac{V_{\mathrm{1\,cav}}}{\Delta}\cdot \kappa_m^{\pm c(s)}, \qquad
\kappa_m^{\pm c(s)}= 
\left(\alpha^{\mathrm{\pm c(s)}}_m + \frac{\beta^{\mathrm{\pm c(s)}}_m}{\omega_\pm^2 L_1^2}\right)\;,
\end{equation}
and $\Delta$ is a distance between borders of two cavities (see Fig. 3), and $\alpha^{\mathrm{\pm c(s)}}_m$ and $\beta^{\mathrm{\pm c(s)}}_m$ are dimensionless geometrical form-factors,
\begin{equation}
\alpha^{\mathrm{\pm c(s)}}_m=  \int\limits_{\mathrm{2\,cav}}\frac{d^3x}{V_2}\,{\cal E}^z_m(\Vec{x})\int\limits_{\mathrm{1\,cav}} \frac{d^3x'}{V_1}\frac{\left|F_\pm(\Vec{x}\,')\right|}{E_0^2}\cdot \frac{\Delta}{|\Vec{x}-\Vec{x}'|}\left\{\begin{array}{c} \cos \left(k_\pm|\Vec{x}-\Vec{x}'|\right) \\ \sin \left(k_\pm|\Vec{x}-\Vec{x}'|\right) \end{array} \right\},    
\end{equation}
\begin{eqnarray}
\beta^{\mathrm{\pm c(s)}}_m & = &  \int\limits_{\mathrm{2\,cav}}\frac{d^3x}{V_2}\,{\cal E}^z_m(\Vec{x})\int\limits_{\mathrm{1\,cav}} \frac{d^3x'}{V_1}\frac{\partial^2_{z'}\left|F_\pm(\Vec{x}\,')\right|}{E_0^2}\cdot \frac{\Delta \cdot L_1^2}{|\Vec{x}-\Vec{x}'|}\left\{\begin{array}{c} \cos \left(k_\pm|\Vec{x}-\Vec{x}'|\right) \\ \sin \left(k_\pm|\Vec{x}-\Vec{x}'|\right) \end{array} \right\} - \notag\\
& - & \int\limits_{\mathrm{2\,cav}} \dfrac{d^3 x}{V_2} \,{\cal E}^z_m(\vec{x})\int\limits_{\mathrm{1\,S}} \dfrac{\rho'd\rho'd\varphi'}{V_1} \dfrac{\partial_{z'}|F_{\pm}(\Vec{x'})|}{E_0^2}\cdot \dfrac{\Delta \cdot L_1^2}{|\Vec{x} - \Vec{x}'|}\left\{\begin{array}{c} \cos \left(k_\pm|\Vec{x}-\Vec{x}'|\right) \\ \sin \left(k_\pm|\Vec{x}-\Vec{x}'|\right) \end{array} \right\} \bigg|_{z' = 0}^{z' = L}\;. 
\label{beta}
\end{eqnarray}
In deriving the expression (\ref{beta})  for simplicity of numerical calculations we transferred the derivative over $z$ to $z'$. The last term of Eq.~(\ref{beta}) appeared as the result of integrating by parts, see Appendix C for details. 

We estimate the number of signal photons for a given mode in receiving 
cavity as follows~\cite{Irastorza:2018dyq},
\begin{equation}
N_s \simeq  \frac{1}{2 \omega} \int\limits_{\mathrm{2\,cav}}d^3x\,\langle |\Vec{E}^{\mathrm{sig}}(\vec{x},t)|^2 \rangle \simeq \frac{V_2}{2 \omega} \langle |E_m^{\mathrm{sig}}(t)|^2\rangle\;.    
\end{equation}
Signal-to-noise ratio has the following form~\cite{Bogorad:2019pbu}, 
\begin{equation}
    \text{SNR} \simeq \frac{N_s}{N_{\mathrm{th}}} \frac{1}{2L_2 Q} \sqrt{\frac{t}{B}}\;,
\end{equation}
where $t$ is a time of measurement, $B$ is a bandwidth of the 
signal, $L_2$ is a receiving cavity length, $N_{\mathrm{th}}=T/\omega$ 
is a number of thermal photons at $T\gg \omega$.  
One has the following limit on coupling,
\begin{equation}
g_{a \gamma \gamma} \simeq \left(\frac{128 \pi^2 TL_2 \Delta^2}{E_0^4 (B^z_c)^2 Q ((\kappa_m^{c+})^2 +(\kappa_m^{s+})^2)V_1^2 V_2} \sqrt{\frac{B}{t}} \text{SNR}\right)^{1/4}\;.    
\label{gaggLinear}
\end{equation}
However, the volume of detecting cavity $V_2=\pi R_2^2\cdot L_2$ is not an independent variable for fixed detection mode. The resonant condition  $\omega_{TM010}=\omega_\pm$
for detection with TM010 mode yields $R_2=x_{01}/\omega_\pm$. Thus, we have
\begin{equation}
\label{gaaa}
g_{a \gamma \gamma} \simeq 2.52\left(\frac{ T \cdot \Delta^2 \,\cdot \, \omega_\pm^2}{E_0^4 (B^z_c)^2 Q ((\kappa_m^{c+})^2 +(\kappa_m^{s+})^2)V_1^2} \sqrt{\frac{B}{t}} \text{SNR}\right)^{1/4}\;. 
\end{equation}
In particular, Eq.~(\ref{gaaa}) 
can be expressed as
\begin{figure}[t]
\centering{\includegraphics[width=0.9\linewidth]{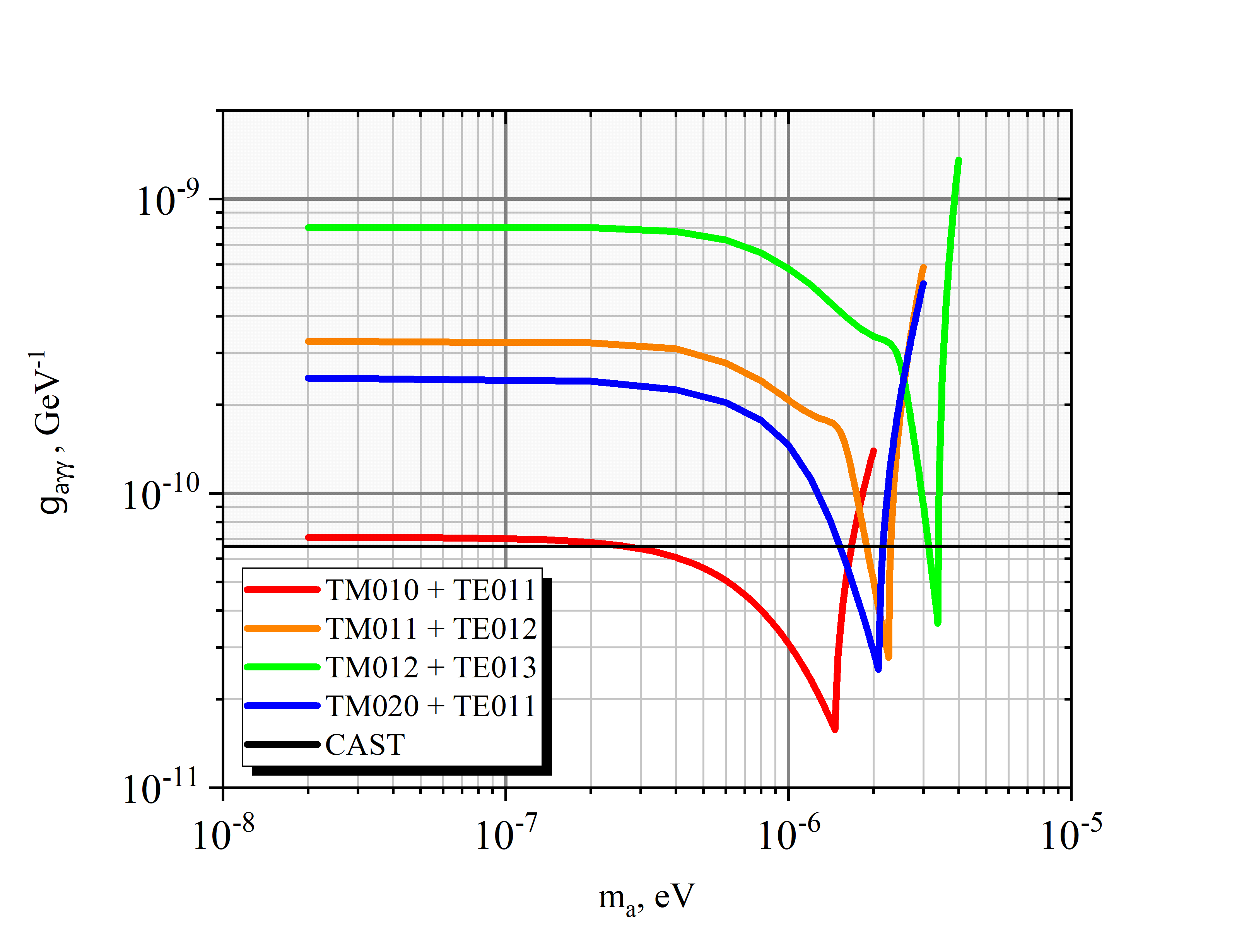}}
\caption{Projected sensitivity of the proposed setup to  ALP mass $m_a$ and $g_{a\gamma\gamma}$ parameter for different sets of pump modes in the production cavity. The black line shows the solar axion constraint from CAST \cite{Anastassopoulos:2017ftl}.
}
\label{fig:Exclusion_plot1}
\end{figure}

\begin{align}
\label{gagg_dependence}
g_{a \gamma \gamma} &\simeq  2.7\cdot 10^{-12}\, \mbox{GeV}^{-1} \left(\frac{T}{1.5\, \mbox{K}}\right)^{1/4} \left(\frac{\Delta}{0.2 \, \mbox{m}}\right)^{1/2}
\left(\frac{\omega_\pm}{1.5\,\cdot \,10^{-6}\, \mbox{eV}}\right)^{1/2}
\left(\frac{V_1}{ 1\, \mbox{m}^3}\right)^{-1/2} \left(\frac{Q}{10^5}\right)^{-1/4}
 \times \notag
\\
&\times\left(\left(\kappa_c^+\right)^2 + \left(\kappa_s^+\right)^2\right)^{-1/4}
\left(\frac{E_0}{0.1\, \mbox{T}}\right)^{-1}
\left(\frac{B_e^z}{10\, \mbox{T}}\right)^{-1/2}
 \left(\frac{t}{10^6\, \mbox{sec}}\right)^{-1/8} \left(\frac{B}{1\, \mbox{Hz}}\right)^{1/8}
\left(\frac{\mbox{SNR}}{5}\right)^{1/4}.
\end{align}
The constraints (\ref{gagg_dependence}) at the parametric plane $(g_{a\gamma\gamma},m_a)$ are shown
in Fig.~\ref{fig:Exclusion_plot1} for different sets of pump modes. The magnitudes of the external parameters,
$T,\,\Delta,\,Q,\,E_0,\,B_e^z,\,t,\,B$ and SNR correspond to their typical values in brackets. 
The resonant constraints on $g_{a\gamma\gamma}$ for larger number of  different pump mode combinations and for different ratio $R_1/L_1$  are shown in Table~1. 


Let us discuss the parametric dependence in Eq.~(\ref{gagg_dependence}). 
First, these bounds are sensitive to the magnitude of the field $E_0$. In particular, factor $0.1$ in the amplitude implies suppression of the bound (\ref{gagg_dependence}) by factor $10$. Second, the dependence on the linear sizes of the setup can be seen from Eq.~(\ref{gaggLinear}). We note that dimensionless form-factors $\kappa_{c(s)}^\pm$ feebly depend on $V_1$ and $V_2$. Therefore, by increasing the linear sizes of the setup by factor of $2$ one can achieve the improved limit on $g_{a\gamma\gamma}$ by factor $2^{3/2}\simeq 2.8$. 



\section{Discussion}\label{sec:discuss}
In the present paper we  have calculated numerically  the time-averaged axion energy 
density $\langle \rho^E \rangle$ 
generated by two electromagnetic modes in the superconducting cylindrical cavity. In 
particular, we have studied
the spatial distribution of $\langle \rho^E \rangle$  for various axion masses and for 
different types of cylindrical  cavity. 
We have shown numerically that there is a parametric resonance if axion produced in nonrelativistic regime for 
the frequency component $\omega_+ = |\omega_1+\omega_2|$  and for certain
combination of pump modes.  
On the contrary, for the frequency component $\omega_-=|\omega_1-\omega_2|$ there is no significant 
resonance in that  regime.

Considering different combinations of pump modes in the production cavity we have shown that they may
have different 
efficiency for axion production. In particular, the highest amplitude for the produced axion field 
came from TE$n_1 p_1 q_1$+TM$n_2 p_2 q_2$ modes with 
$n_1=n_2$ and even $q_1+q_2$. Concerning 
different radius-to-length ratio $R/L$ for the 
cavity, we have shown that for a ``pancake-like'' cavity, $L/R \lesssim 1$, the maximum ALP energy 
density outside the  cavity  is produced along the cylinder axis $z$. 
Otherwise, for ``prolonged'' design of the cavity,  
$L/R \gg 1$, the maximum energy density outside the cylinder is 
produced near the side surface (this case was studied in Ref.~\cite{Janish:2019dpr}).

We also discuss the detection of ALPs in the additional 
separated cavity, which is filled with the strong magnetic field.
We estimated the projected sensitivity of the proposed setup on
the ALPs parameter space $(m_a,g_{a\gamma\gamma})$. 
We have shown that the best constraint came from the mode combination TM010+TE011 in the production cavity. 

In addition, we have calculated the sensitivity curves
for high order pump modes of the production cavity. 
The  advantage of generating high order pump modes  is that  one can
probe the region with higher masses of ALPs.
However the peak sensitivity to the ALPs coupling is decreased in 
that case. Although the resonances at 
Fig.~\ref{fig:Exclusion_plot1} are quite narrow, we can test 
larger area of parameters by exciting 
different set of high order pump modes. Nevertheless, for fixed cavity geometry
there are still spaces between resonant peaks at the exclusion 
plot.  One can modify eigenmodes by adjusting the geometry of given cavity. Therefore, the
relevant  unconstrained area can be probed. 


Note that our setup can be easily generalized to 
other designs of the detection cavity. In particular,  the analysis 
can be easily applied the toroidal detection cavity
\cite{Janish:2019dpr}. 
By exciting high order pump modes one can shift the resonant
bounds of \cite{Janish:2019dpr} to the area of higher ALP masses.
Another interesting task is to consider parametric resonances 
for the single cavity which produces and detects 
ALPs~\cite{Bogorad:2019pbu}. I addition, it is instructive to probe
hidden photon~\cite{Kim:2020ask} in SRF cavity.
We leave these tasks for further work.






\paragraph{Acknowledgements} The authors thank Sergey Troitsky, Dmitry Levkov, Alexander Panin, Ilya Kopchinskii, Yuri Senichev, Leonid Kravchuk, Valentin Paramonov, Andrey Egorov,  Leonid Kuzmin and Andrey Pankratov for helpful discussions. Numerical calculations were performed on the Computational Cluster of the Theoretical division of INR RAS. The work of P.S. and M.F. was supported by RSF grant 18-12-00258.

\appendix
\section{Retarded Green function}\label{app}
In this Appendix we show some details of deriving Eqs.~(\ref{GretTileSup2}), (\ref{GretTileSup3}). The Klein-Gordon's retarded Green function can be written as follows \cite{Polyanin},
\begin{equation}
G_{\mathrm{ret}}(\vec{x},\vec{x}',t,t') = \frac{1}{4\pi}\Bigl(\frac{\delta(t-t'-R)}{R} - \theta(t-t' -R) \frac{ m_a}{u} J_{1}(m_a  u)\Bigr)\;, \label{ALPRetGrF}
\end{equation}
where $u \equiv \sqrt{(t-t')^2-R^2}$ and $R\equiv |\vec{x}-\vec{x}'|$. The first term in Eq.~(\ref{ALPRetGrF})  describes the retarded Green function of the massless scalar field, so that the mass dependence is only in the second term. In Eq.~(\ref{Green}) the retarded Green function should be integrated over $t'$ with the electromagnetic invariant $\vec{E}(x',t')\cdot \vec{B}(x',t')$. For considered modes $t'$ dependence decouples as $e^{-i\omega t'}$. For purpose of integration of the second term of Eq.~(\ref{ALPRetGrF}) the textbook integral  \cite{Ryzhik} can be used,
\begin{equation}
\int\limits_R^\infty d\tilde{t}\,\frac{\mathrm{e}^{i\omega \tilde{t}}}{\sqrt{\tilde{t}^2-R^2}}
J_1(m_a \sqrt{\tilde{t}^2-R^2}) = \frac{1}{m_a R} 
\left(\mathrm{e}^{iR\omega} - \mathrm{e}^{iR\sqrt{\omega^2-m_a^2}}\right)\;.
\end{equation}
Integrating the Green function (\ref{ALPRetGrF}), one obtains 
\begin{equation}
\int\limits^{t-R}_{-\infty} dt'\, G_{\mathrm{ret}}(\vec{x},\vec{x}',t,t')\,\mathrm{e}^{-i\omega t'}= \frac{1}{4\pi R}\,\mathrm{e}^{-i\omega t+i R k}\;, \label{GretTileOscill3}
\end{equation}
where $k=\sqrt{\omega^2-m_a^2}$. If $m_a<\omega$, $ik$ should be replaced by $\kappa=\sqrt{m_a^2-\omega^2}$. As a result, one immediately obtains Eqs.~(\ref{GretTileSup2}), (\ref{GretTileSup3}).  

\section{Exact form of $TM_{npq}$ and $TE_{npq}$ modes}\label{AppB}
$TM_{npq}$ and $TE_{npq}$ modes are expressed in terms of the electric and magnetic fields \cite{Hill} as
\begin{align}
{\cal E}^{TMnpq}_z & =  {\cal E}_0J_{n}\left(\dfrac{x_{np}}{R}\rho\right)\left\{\begin{array}{c}
\sin n\varphi \\
 \cos n\varphi
\end{array} \right\}\cos \left(\dfrac{q\pi}{L}z\right), \\
{\cal E}^{TMnpq}_{\rho} & = \dfrac{-{\cal E}_0}{k^2_{npq} - (q\pi/L)^2}\cdot\dfrac{q\pi}{L}\cdot \dfrac{x_{np}}{R}\cdot J'_{n}\left(\dfrac{x_{np}}{R}\rho \right)\left\{\begin{array}{c}
\sin n\varphi \\
\cos n\varphi
\end{array} \right\}\sin \left(\dfrac{q\pi}{L}z\right), \\
{\cal E}^{TMnpq}_{\varphi} & =  \dfrac{-{\cal E}_0}{k^2_{npq} - (q\pi/L)^2}\cdot\dfrac{1}{\rho}\cdot \dfrac{nq\pi}{L}\cdot J_{n}\left(\dfrac{x_{np}}{R}\rho \right)\left\{\begin{array}{c}
\cos n\varphi \\
-\sin n\varphi
\end{array} \right\}\sin \left(\dfrac{q\pi}{L}z\right), \\
{\cal B}^{TMnpq}_{z} & = 0, \\
{\cal B}^{TMnpq}_{\rho} & = \dfrac{-i\omega^{TM}_{npq}{\cal E}_0}{k^2_{npq} - (q\pi/L)^2}\cdot \dfrac{n}{\rho}\cdot J_{n}\left(\dfrac{x_{np}}{R}\rho \right)\left\{\begin{array}{c}
\cos n\varphi \\
-\sin n\varphi
\end{array} \right\}\cos \left(\dfrac{q\pi}{L}z\right), \\
{\cal B}^{TMnpq}_{\varphi} & =  \dfrac{i\omega^{TM}_{npq}{\cal E}_0}{k^2_{npq} - (q\pi/L)^2}\cdot\dfrac{x_{np}}{R}\cdot J'_{n}\left(\dfrac{x_{np}}{R}\rho \right)\left\{\begin{array}{c}
\sin n\varphi \\
\cos n\varphi
\end{array} \right\}\cos \left(\dfrac{q\pi}{L}z\right);
\end{align}

\begin{align}
{\cal B}^{TEnpq}_z & =  {\cal B}_0J_{n}\left(\dfrac{x'_{np}}{R}\rho\right)\left\{\begin{array}{c}
\sin n\varphi \\
 \cos n\varphi
\end{array} \right\}\sin \left(\dfrac{q\pi}{L}z\right), \\
{\cal B}^{TEnpq}_{\rho} & = \dfrac{{\cal B}_0}{k^2_{npq} - (q\pi/L)^2}\cdot\dfrac{q\pi}{L}\cdot \dfrac{x'_{np}}{R}\cdot J'_{n}\left(\dfrac{x'_{np}}{R}\rho \right)\left\{\begin{array}{c}
\sin n\varphi \\
\cos n\varphi
\end{array} \right\}\cos \left(\dfrac{q\pi}{L}z\right), \\
{\cal B}^{TEnpq}_{\varphi} & =  \dfrac{{\cal B}_0}{k^2_{npq} - (q\pi/L)^2}\cdot\dfrac{1}{\rho}\cdot \dfrac{nq\pi}{L}\cdot J_{n}\left(\dfrac{x'_{np}}{R}\rho \right)\left\{\begin{array}{c}
\cos n\varphi \\
-\sin n\varphi
\end{array} \right\}\cos \left(\dfrac{q\pi}{L}z\right), \\
{\cal E}^{TEnpq}_{z} & = 0, \\
{\cal E}^{TEnpq}_{\rho} & = \dfrac{i\omega^{TE}_{npq}{\cal B}_0}{k^2_{npq} - (q\pi/L)^2}\cdot \dfrac{n}{\rho}\cdot J_{n}\left(\dfrac{x'_{np}}{R}\rho \right)\left\{\begin{array}{c}
\cos n\varphi \\
-\sin n\varphi
\end{array} \right\}\sin \left(\dfrac{q\pi}{L}z\right), \\
{\cal E}^{TEnpq}_{\varphi} & =  \dfrac{-i\omega^{TE}_{npq}{\cal B}_0}{k^2_{npq} - (q\pi/L)^2}\cdot\dfrac{x'_{np}}{R}\cdot J'_{n}\left(\dfrac{x'_{np}}{R}\rho \right)\left\{\begin{array}{c}
\sin n\varphi \\
\cos n\varphi
\end{array} \right\}\sin \left(\dfrac{q\pi}{L}z\right).
\end{align}

\section{The gradient of axion field}
For the case $m_a < \omega_{\pm}$ let us calculate the axion field's gradient (calculation for $m_a > \omega_{\pm}$ is similar). Using Eq.~\eqref{GretTileSup2} and the relation $\vec{\nabla}_{x} \, f(|\vec{x} - \vec{x}'|) = - \vec{\nabla}_{x'} \, f(|\vec{x} - \vec{x}'|)$ one finds
\begin{equation}
\vec{\nabla} a_\pm(\Vec{x},t) = -g_{a\gamma\gamma} \; \Re \mathrm{e}\,\int\limits_{V_{\mathrm{cav}}} d^3x'\,\frac{F_\pm(\Vec{x}')}{4\pi} \left\{ - \vec{\nabla}_{x'} \dfrac{\mathrm{e}^{-i\omega_\pm t + i|\Vec{x}-\Vec{x}'|k_\pm }}{|\vec{x} - \vec{x}'|} \right\} \;.
\end{equation}



Next, integrating this equation by parts and using Stokes' theorem one gets
\begin{eqnarray}
\vec{\nabla} a_\pm(\Vec{x},t) & = & - g_{a\gamma\gamma} \; \Re \mathrm{e}\,\int\limits_{V_{\mathrm{cav}}} d^3x'\,\frac{ \vec{\nabla}_{x'}F_{\pm}(\vec{x}')}{4\pi|\vec{x} - \vec{x}'|} \mathrm{e}^{-i\omega_\pm t + i|\Vec{x}-\Vec{x}'|k_\pm } \; + \notag\\
& & +  g_{a\gamma\gamma} \; \Re \mathrm{e}\,\oint\limits_{\partial V_{\mathrm{cav}}} \vec{n}\,d\sigma'\, \dfrac{F_{\pm}(\vec{x}')}{4\pi} \cdot \dfrac{\mathrm{e}^{-i\omega_\pm t + i|\Vec{x}-\Vec{x}'|k_\pm }}{|\vec{x} - \vec{x}'|} \;.
\end{eqnarray}
Direct substitution yields the condition 
$F_{\pm}(\vec{x}') = 0 $ on the cavity walls. 
Therefore, the second term in the last formula is zero and we get
\begin{equation}
\vec{\nabla} a_\pm(\Vec{x},t) =  - g_{a\gamma\gamma} \; \Re \mathrm{e}\,\int\limits_{V_{\mathrm{cav}}} d^3x'\,\frac{ \vec{\nabla}_{x'}F_{\pm}(\vec{x}')}{4\pi|\vec{x} - \vec{x}'|} \mathrm{e}^{-i\omega_\pm t + i|\Vec{x}-\Vec{x}'|k_\pm } \;. 
\end{equation}





Let us derive the expression $\vec{\mathcal{E}}_m \cdot \vec{\nabla}\,(\vec{B}_e(\Vec{x})\cdot\vec{\nabla}\,a(\Vec{x},t))$ given the conditions $\vec{B}_e = B^z_e \cdot \vec{e}_z$, $B^z_e = \mathrm{const}$, and $\vec{\mathcal{E}}_m(\vec{x}) = \mathcal{E}^z_m(\vec{x}) \cdot \vec{e}_z$. Integrating by parts and taking into account that $\partial_{z'} F_{\pm}(\vec{x}')\neq 0$ on the cavity wall we arrive at the formula
\begin{eqnarray}
\vec{\mathcal{E}}_m \cdot \vec{\nabla}(\vec{B}_e \cdot \vec{\nabla} a_\pm(\Vec{x},t))  =   - g_{a\gamma\gamma} B^z_e \mathcal{E}^z_m(\vec{x}) \; \Re \mathrm{e}\,\int\limits_{V_{\mathrm{cav}}} d^3x'\,\frac{ \partial^2_{z'}F_{\pm}(\vec{x}')}{4\pi|\vec{x} - \vec{x}'|} \mathrm{e}^{-i\omega_\pm t + i|\Vec{x}-\Vec{x}'|k_\pm } \; + \notag\\
  +  g_{a\gamma\gamma} B^z_e \mathcal{E}^m_z(\vec{x}) \; \Re \mathrm{e}\,\int\limits_{\partial V_{\mathrm{cav}}} \rho'd\rho'd\varphi' \,\frac{ \partial_{z'}F_{\pm}(\vec{x}')}{4\pi|\vec{x} - \vec{x}'|} \mathrm{e}^{-i\omega_\pm t + i|\Vec{x}-\Vec{x}'|k_\pm }\bigg|_{z' = 0}^{z' = L}\;.
\end{eqnarray}
Taking out dimensionful values, one comes to (\ref{beta}).







\begin{figure}[h!]
\begin{minipage}[h]{0.49\linewidth}
\center{\includegraphics[width=1\linewidth]{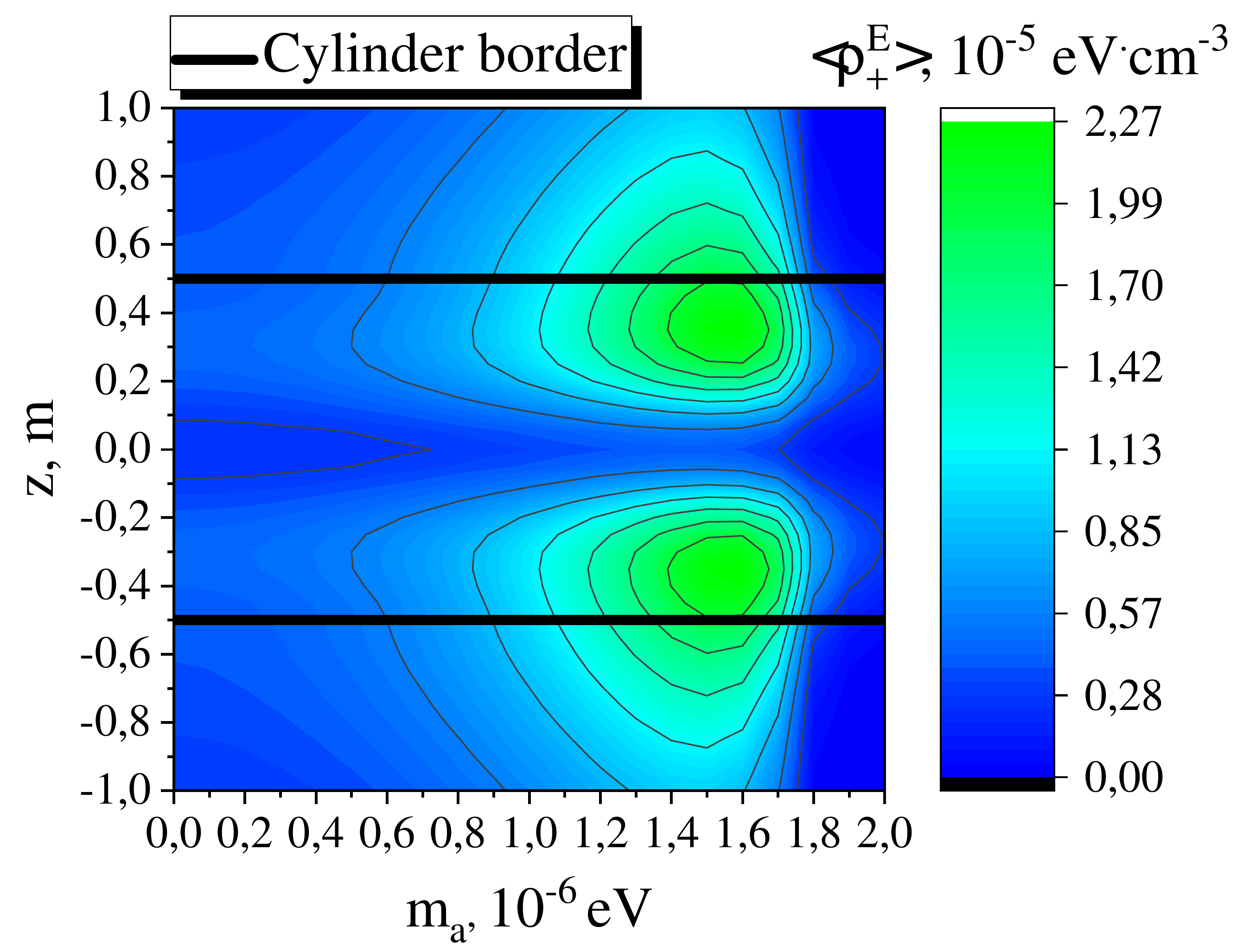}}
\center{\includegraphics[width=1\linewidth]{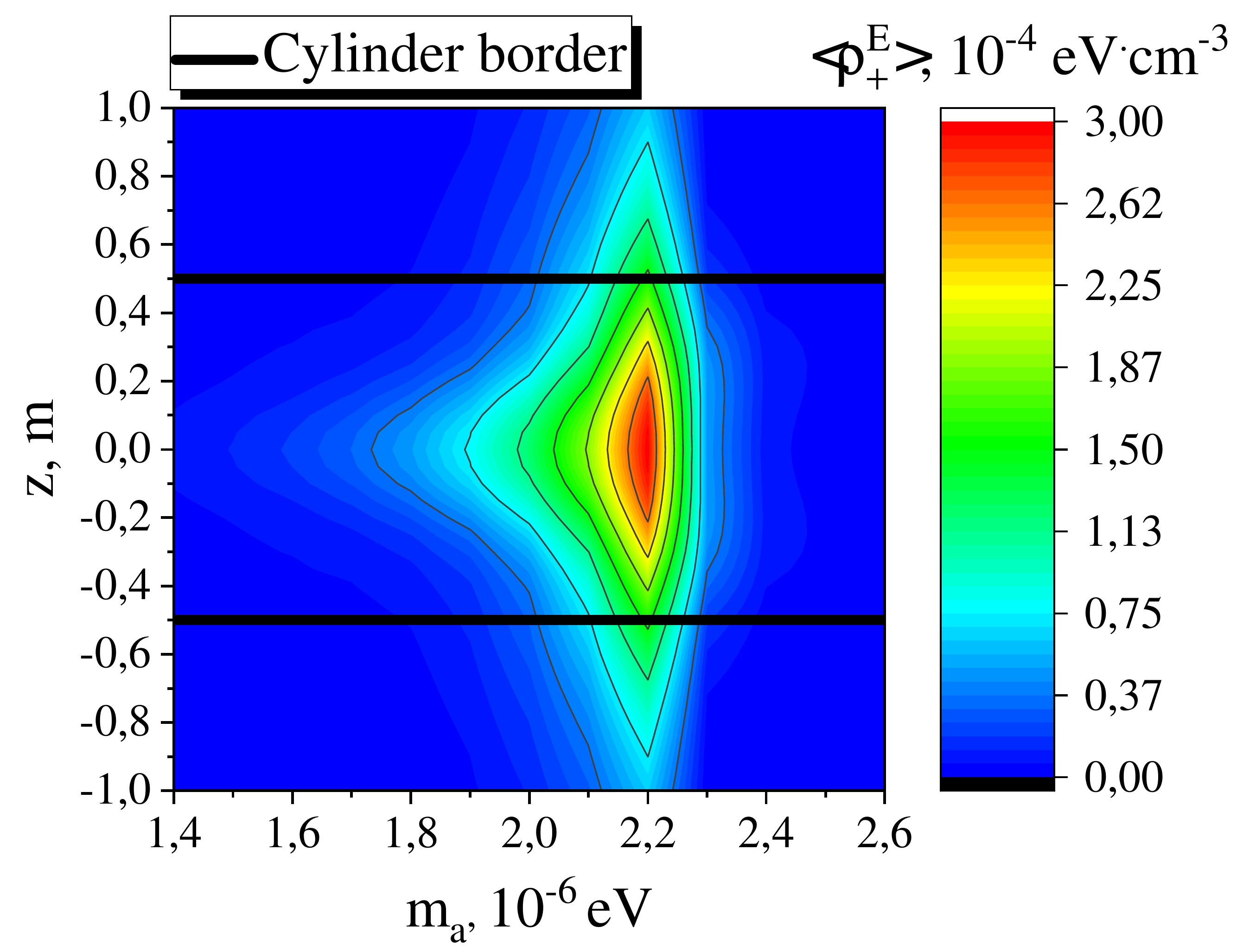}}
\center{\includegraphics[width=1\linewidth]{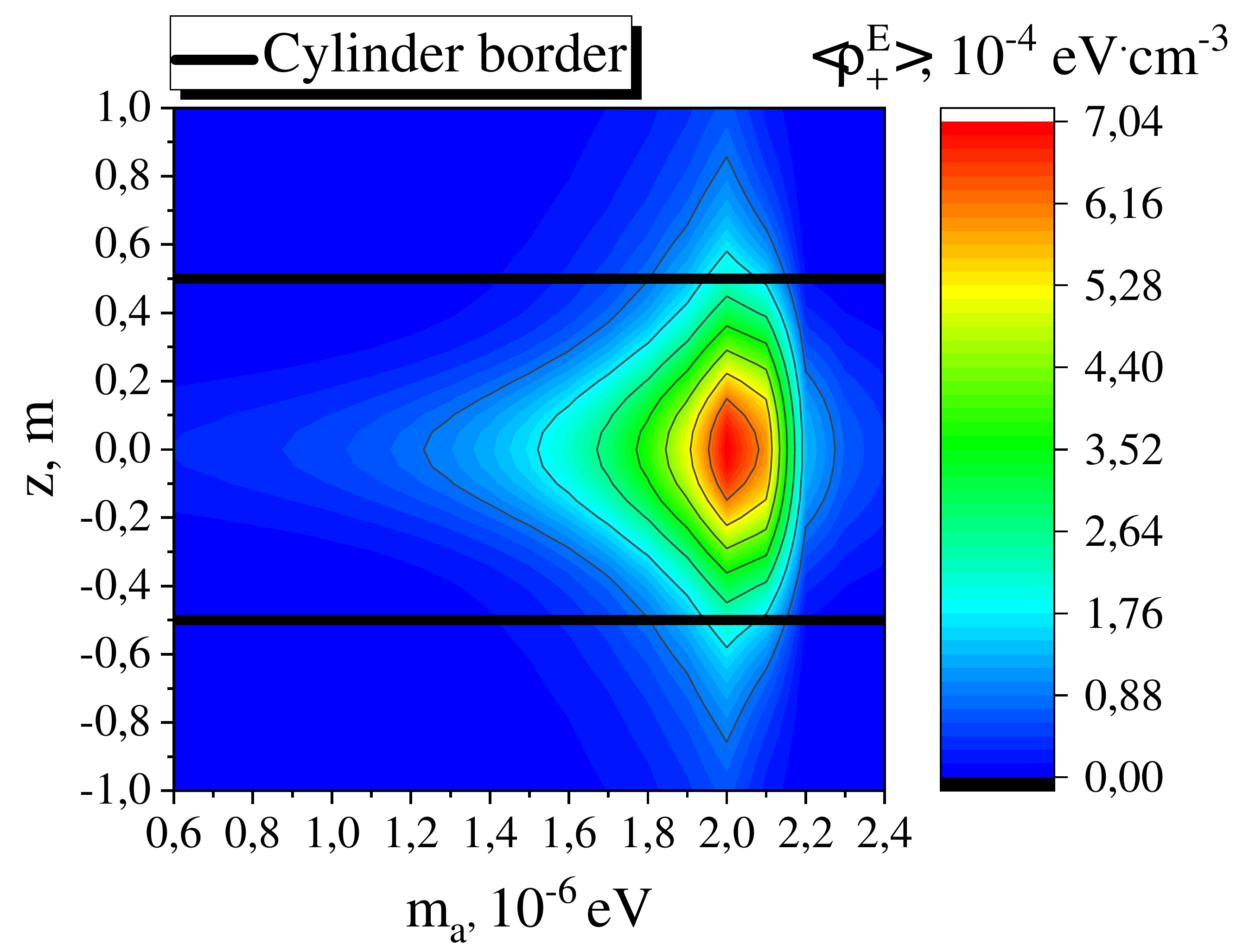}}
\end{minipage}
\hfill
\begin{minipage}[h]{0.49\linewidth}
\center{\includegraphics[width=1\linewidth]{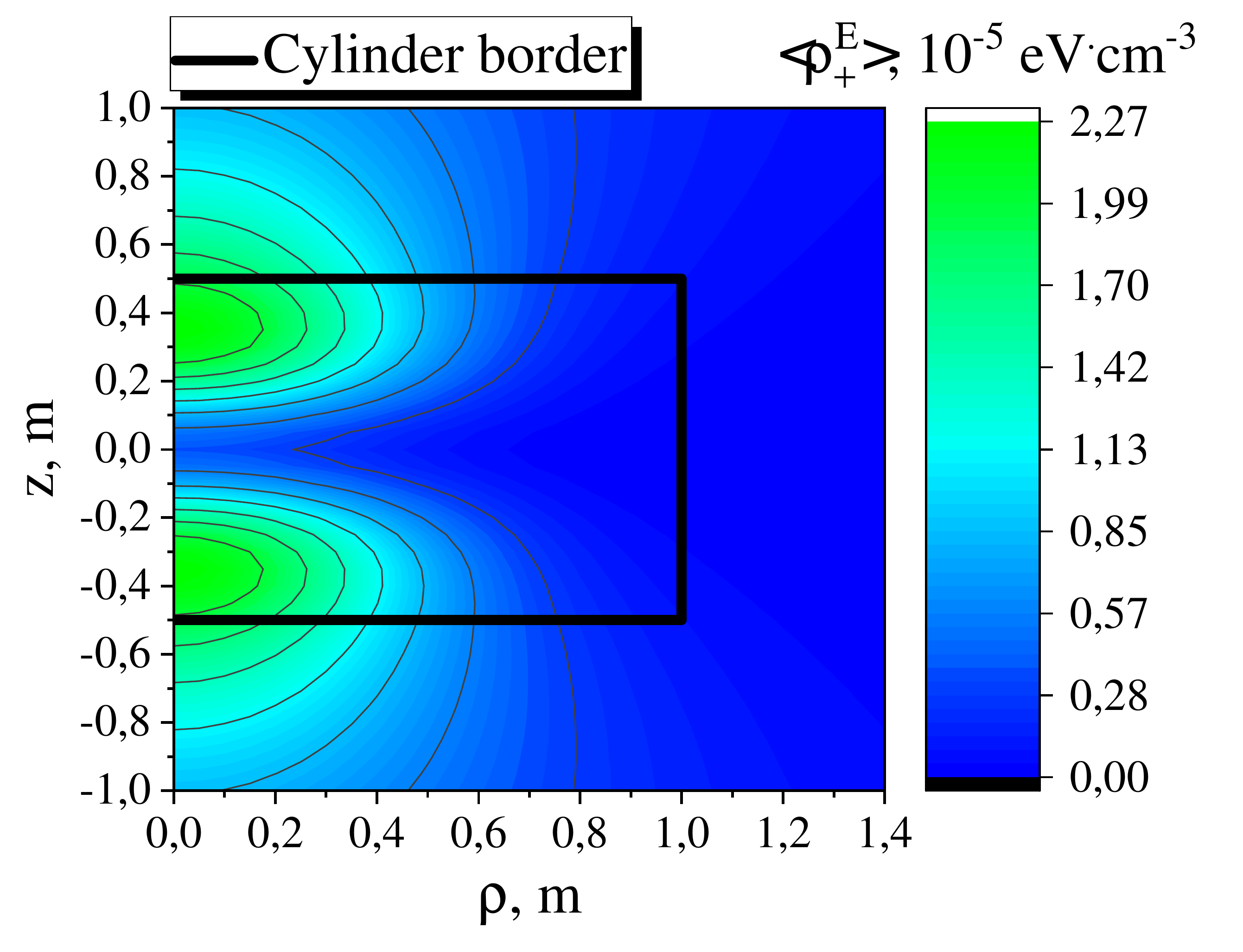}}
\center{\includegraphics[width=1\linewidth]{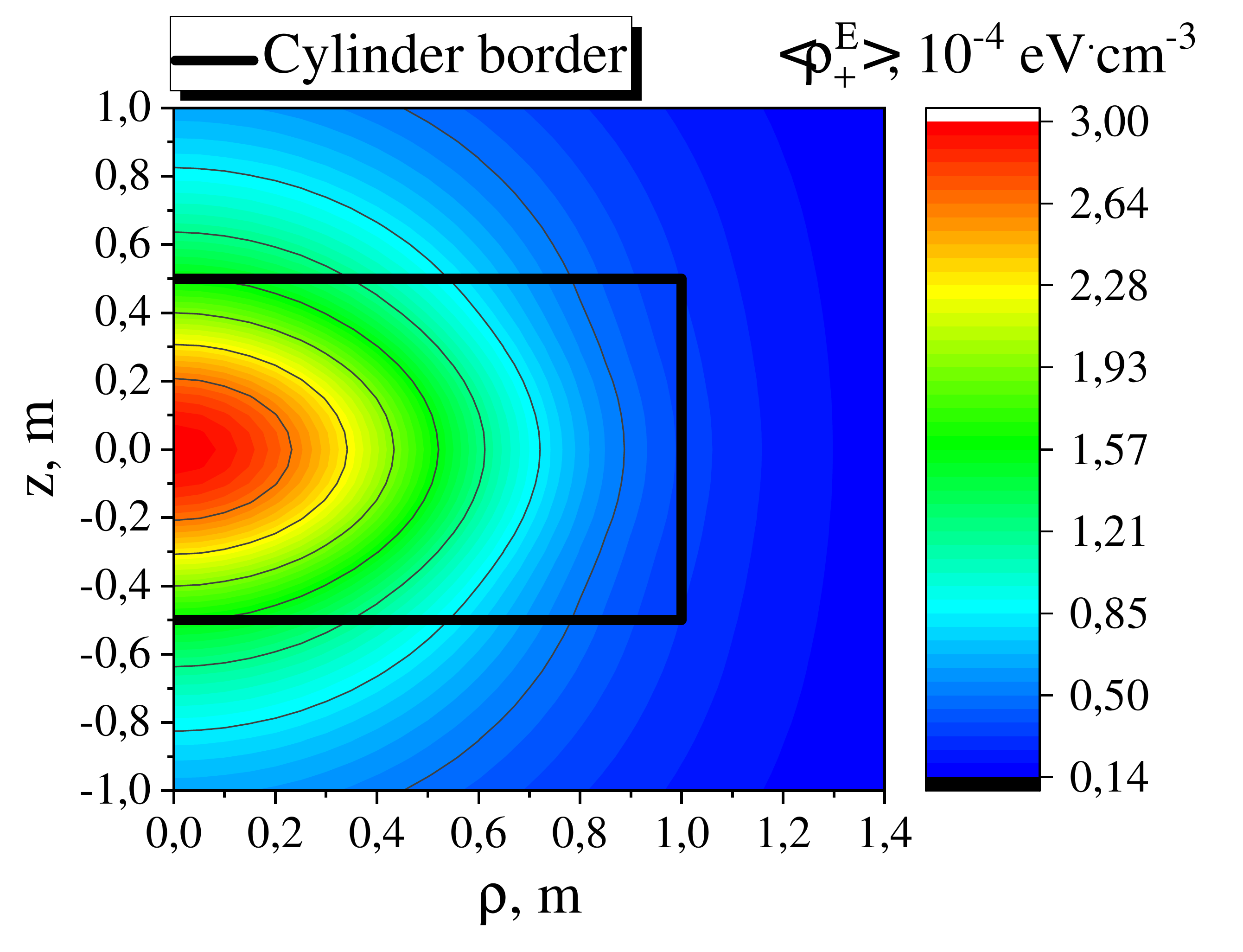}}
\center{\includegraphics[width=1\linewidth]{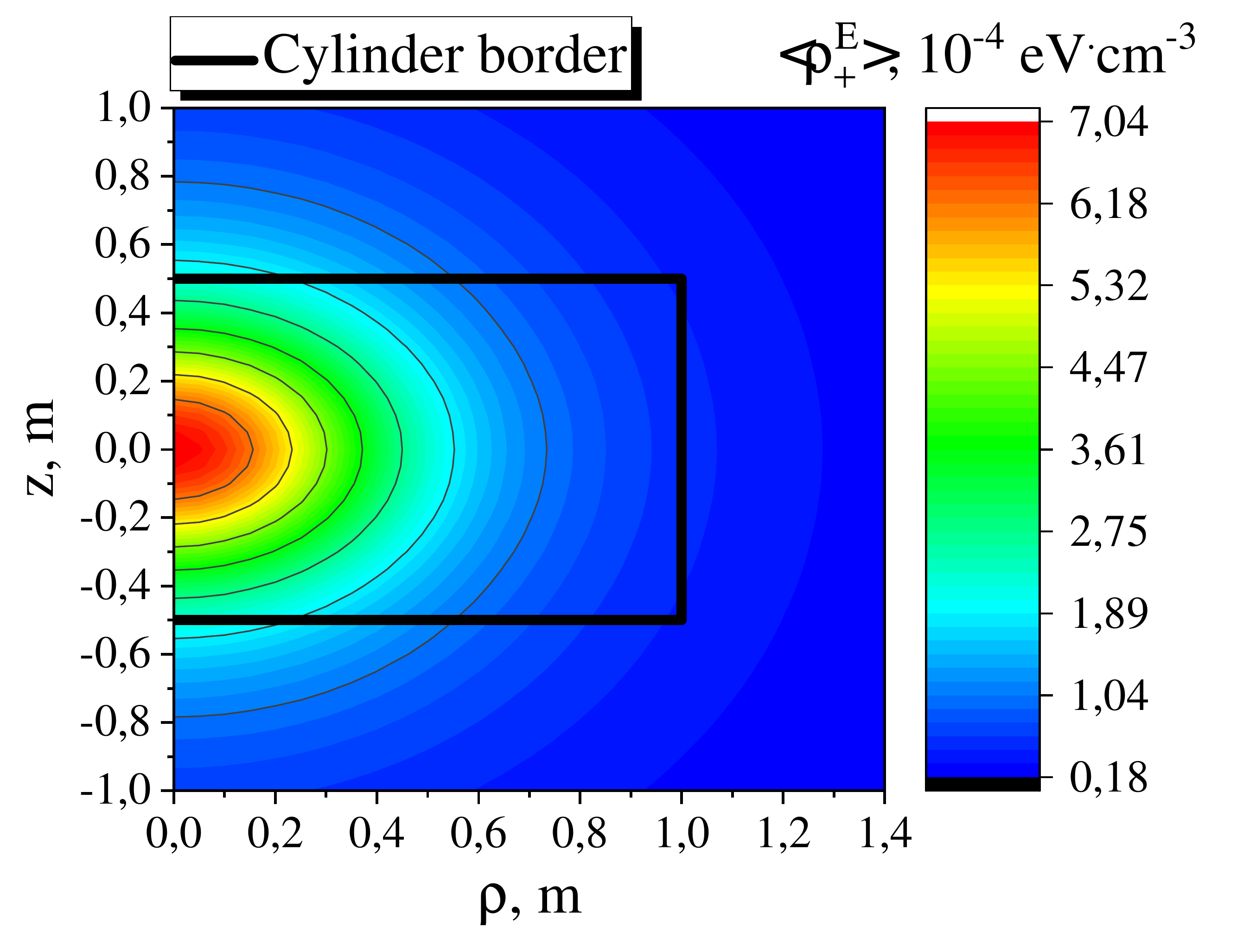}}
\end{minipage}
\caption{Contour plot for the time-averaged energy density $\braket{\rho^E_{+}}$ evaluated on the cylinder axis as function of axion mass $m_a$ and the distance $z$ from center of cavity (left panels) and spatial distribution of $\braket{\rho^E_+}$ on the cavity section along its axis $(\rho,\,z)$ (right panels) with TM011+TE011 $\omega_+=17.8\cdot 10^{-7}$ eV (top), TM011+TE012 $\omega_+=22.6\cdot 10^{-7}$ eV (middle), and TM020+TE011 $\omega_+=21.0\cdot 10^{-7}$ eV (bottom) pump modes. Cavity dimensions: $L = 1$ m, $R = 1$ m.}\label{fig:C1}
\end{figure}

\begin{figure}[h!]
\begin{minipage}[h]{0.32\linewidth}
\center{\includegraphics[width=1\linewidth]{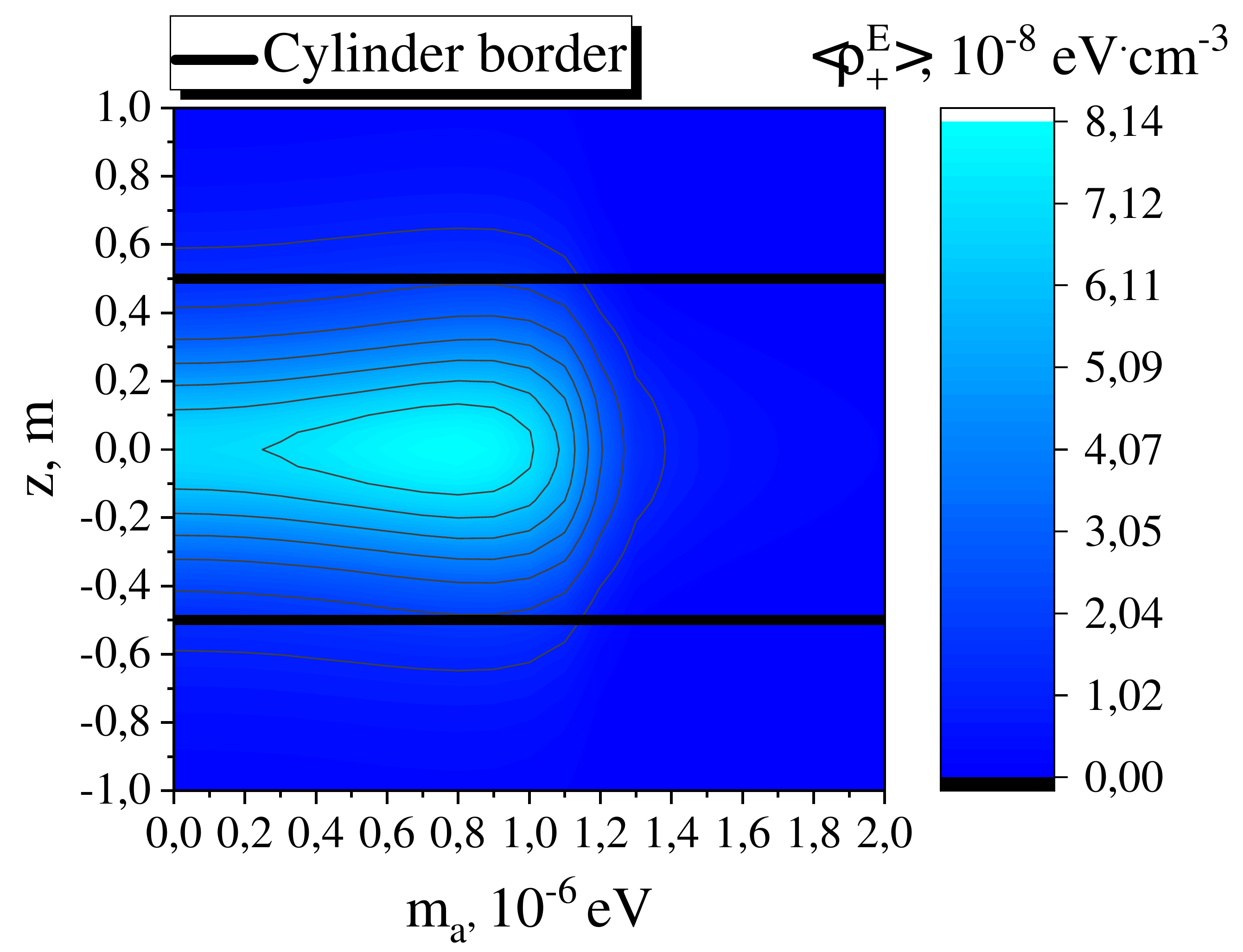}}
\center{\includegraphics[width=1\linewidth]{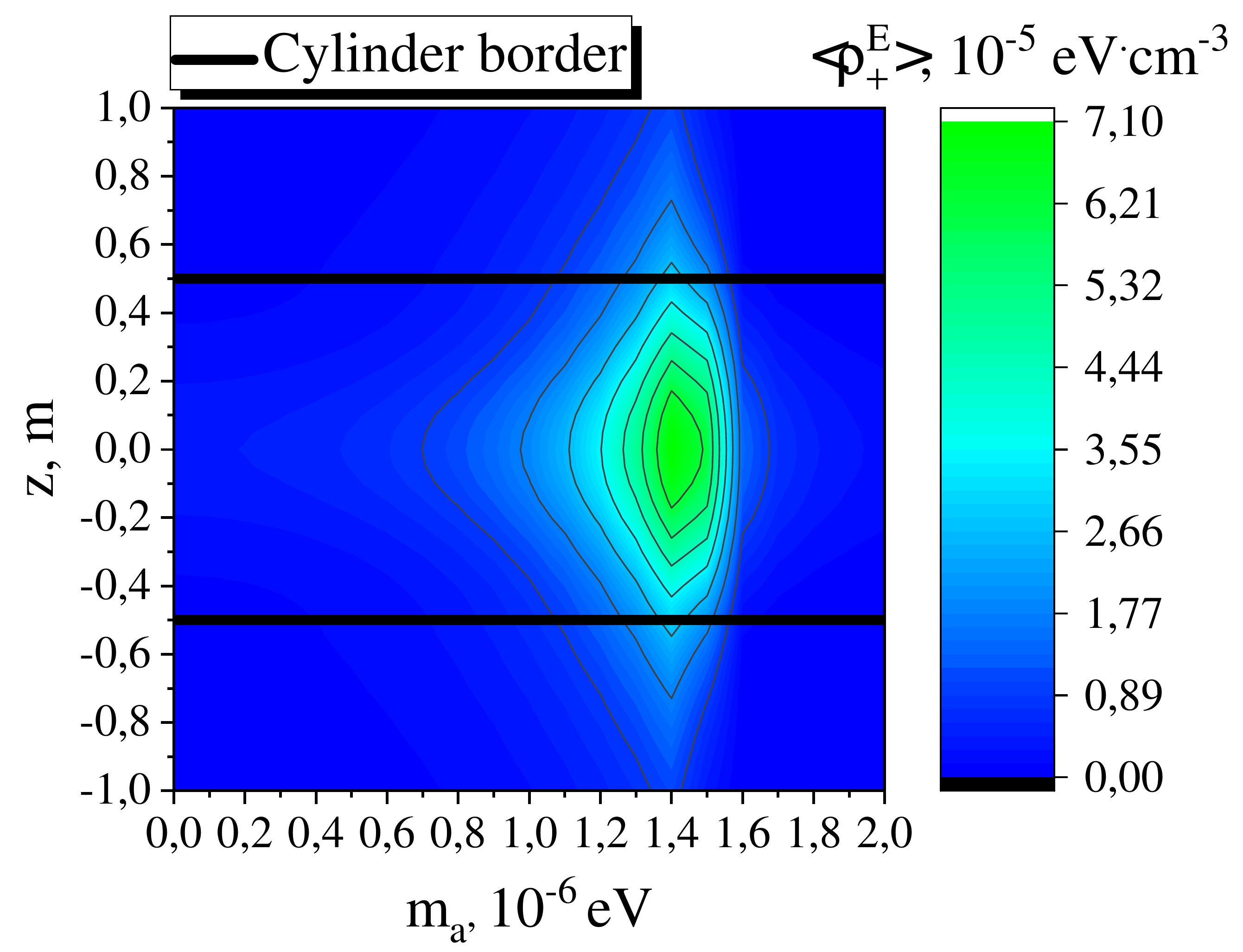}}
\hfill
\end{minipage}
\begin{minipage}[h]{0.32\linewidth}
\center{\includegraphics[width=1\linewidth]{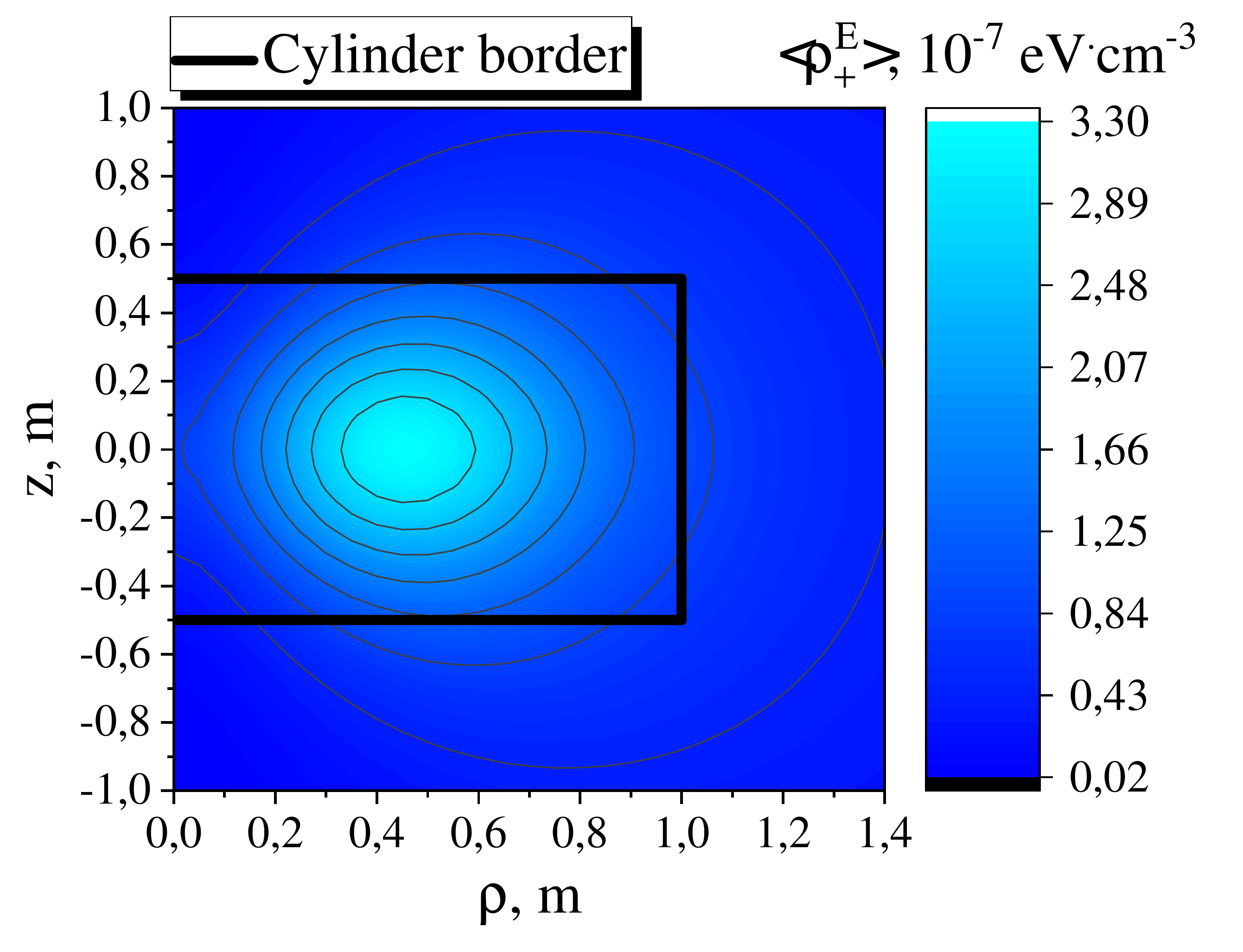}}
\center{\includegraphics[width=1\linewidth]{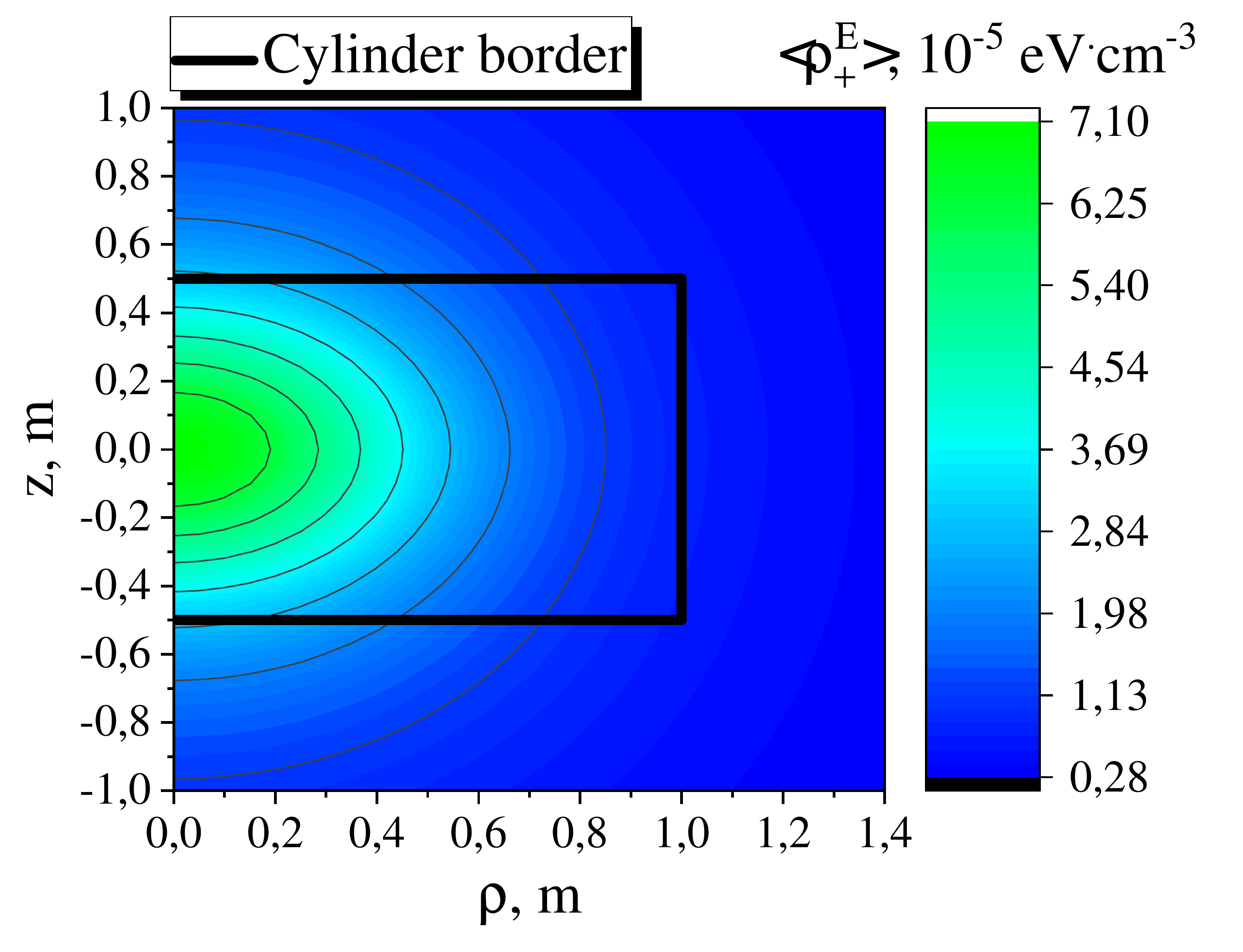}}
\end{minipage}
\begin{minipage}[h]{0.32\linewidth}
\center{\includegraphics[width=1\linewidth]{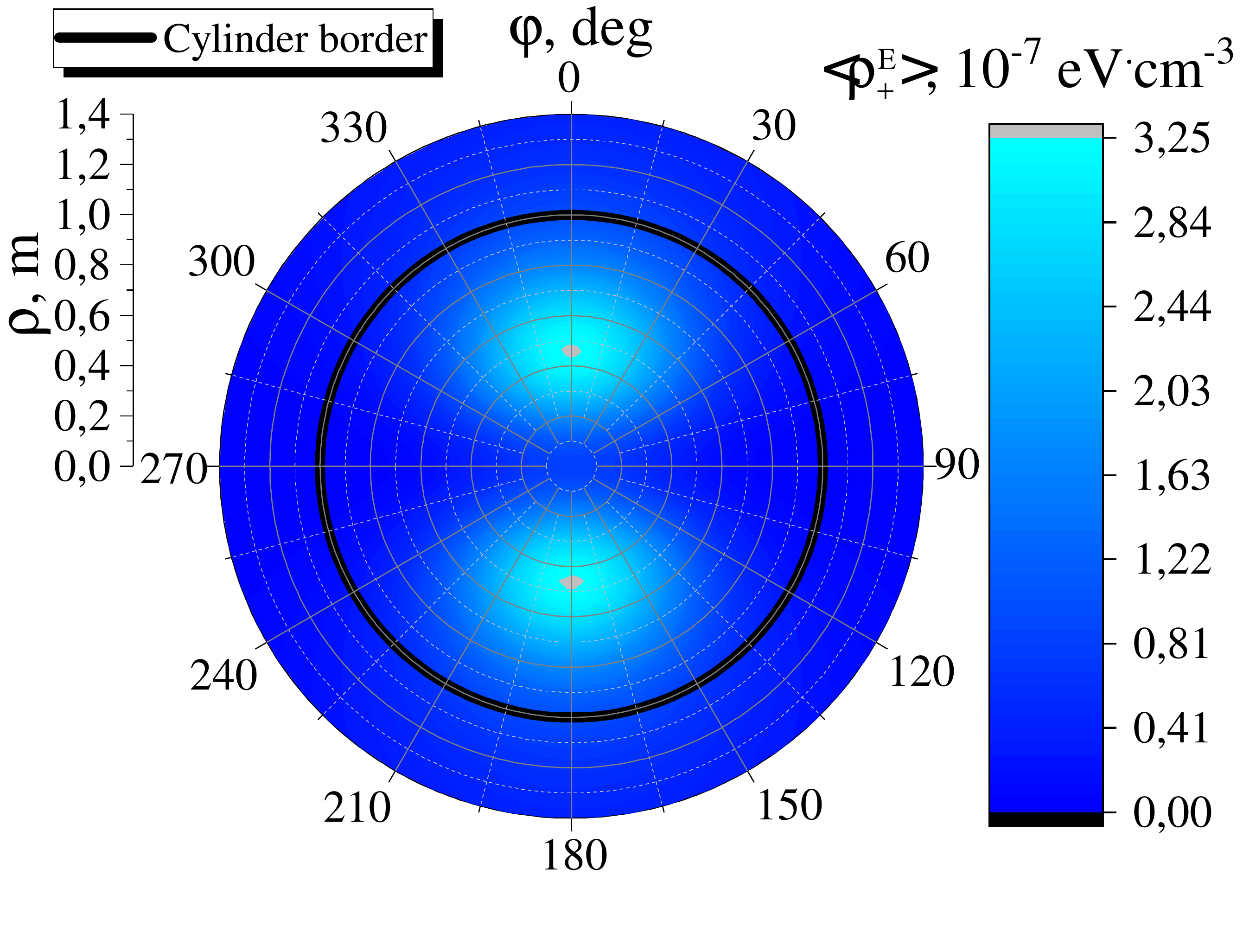}}
\center{\includegraphics[width=1\linewidth]{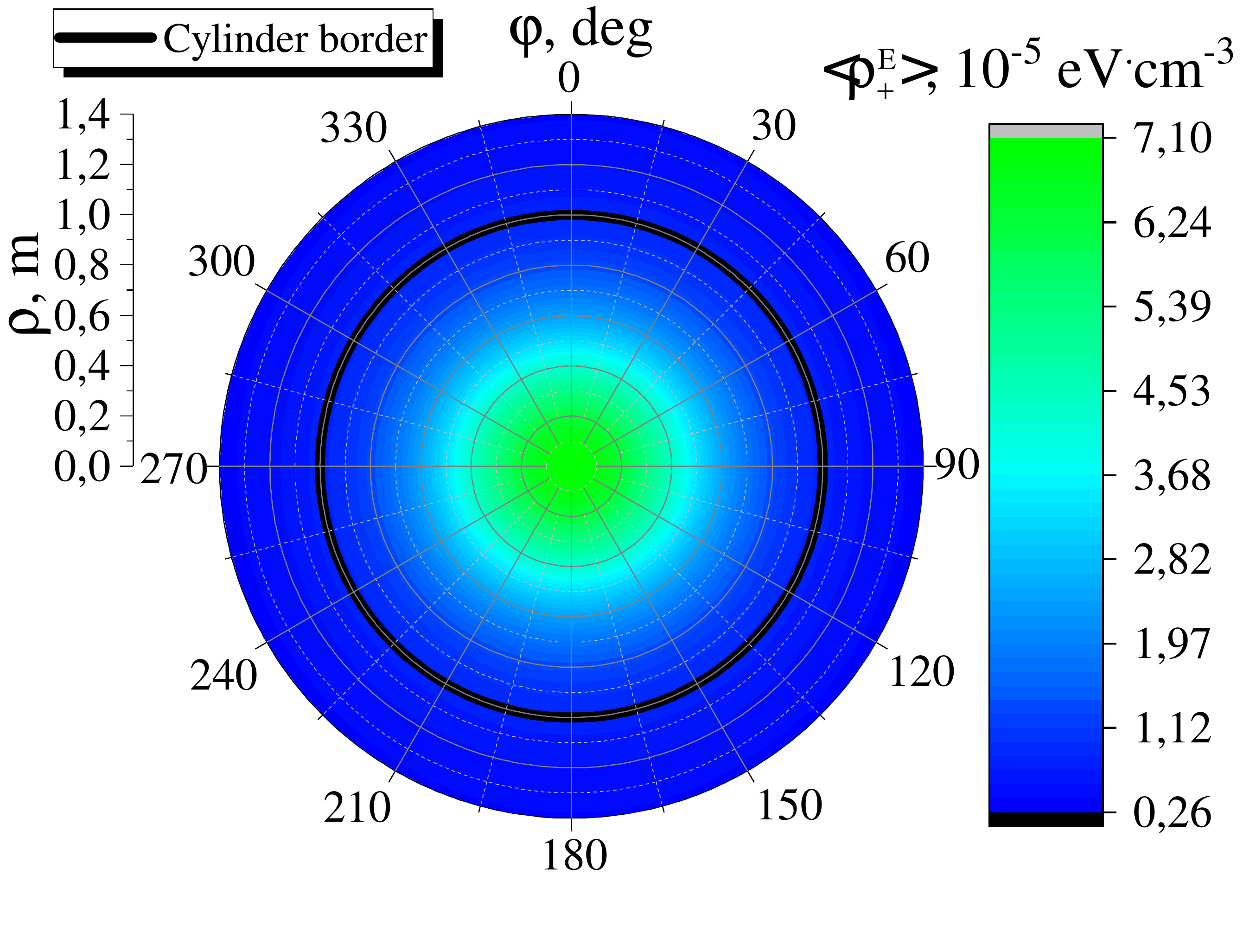}}
\end{minipage}
\caption{Contour plot for the time-averaged energy density $\braket{\rho^E_{+}}$ evaluated on the cylinder axis as function of axion mass $m_a$ and the distance $z$ from center of cavity (left panels) and spatial distribution of $\braket{\rho^E_+}$ on the cavity section along its axis $(\rho,\,z)$ (middle panels) and $(\rho,\,\varphi)$ (right panels) with TM010+TE111 $\omega_+=12.1\cdot 10^{-7}$eV (top), TM110+TE111 $\omega_+=14.9\cdot 10^{-7}$ eV (bottom) pump modes. Cavity dimensions: $L = 1$ m, $R = 1$ m.}\label{fig:C1}
\end{figure}



\begin{figure}[h!]
\begin{minipage}[h]{0.49\linewidth}
\center{\includegraphics[width=1\linewidth]{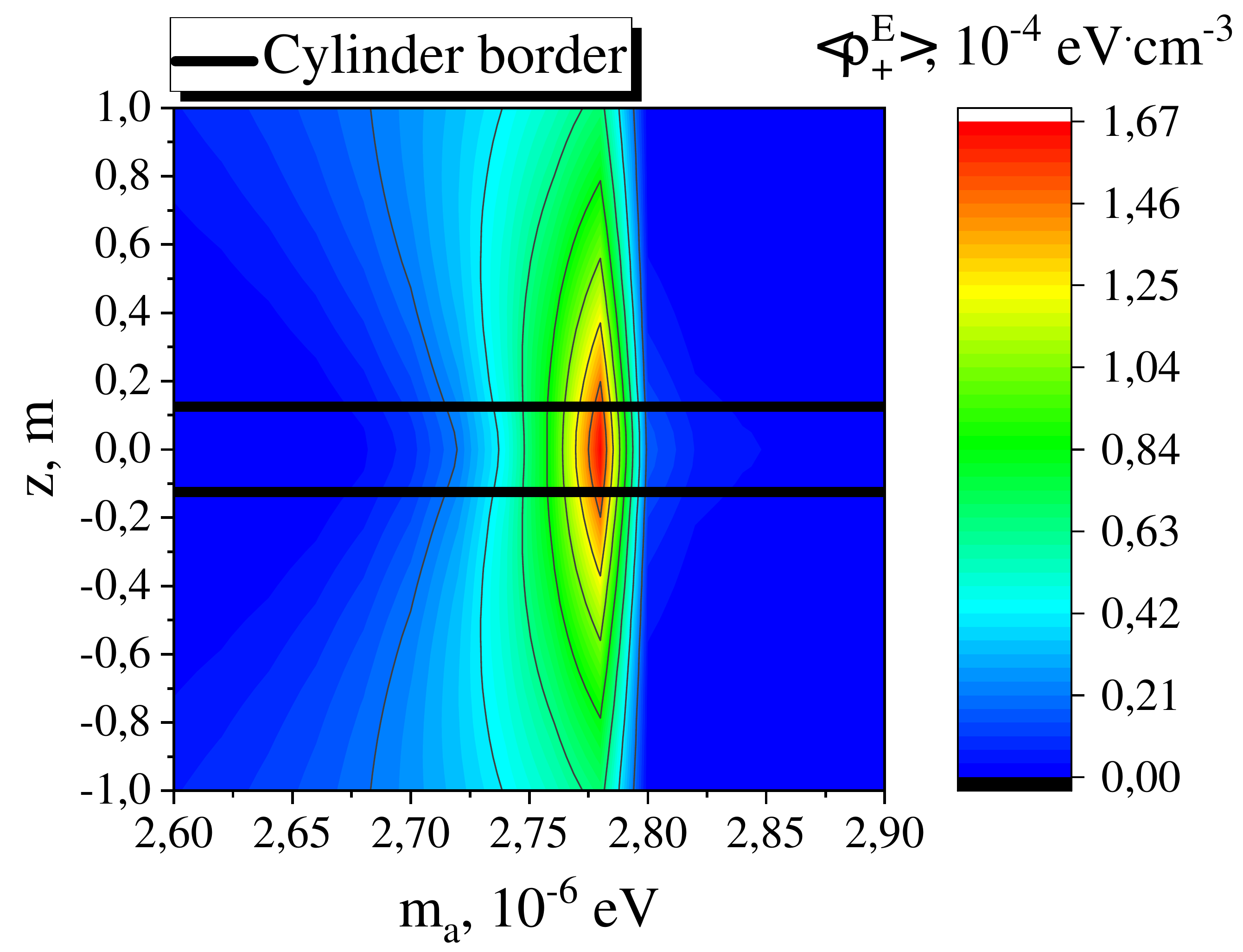}}
\center{\includegraphics[width=1\linewidth]{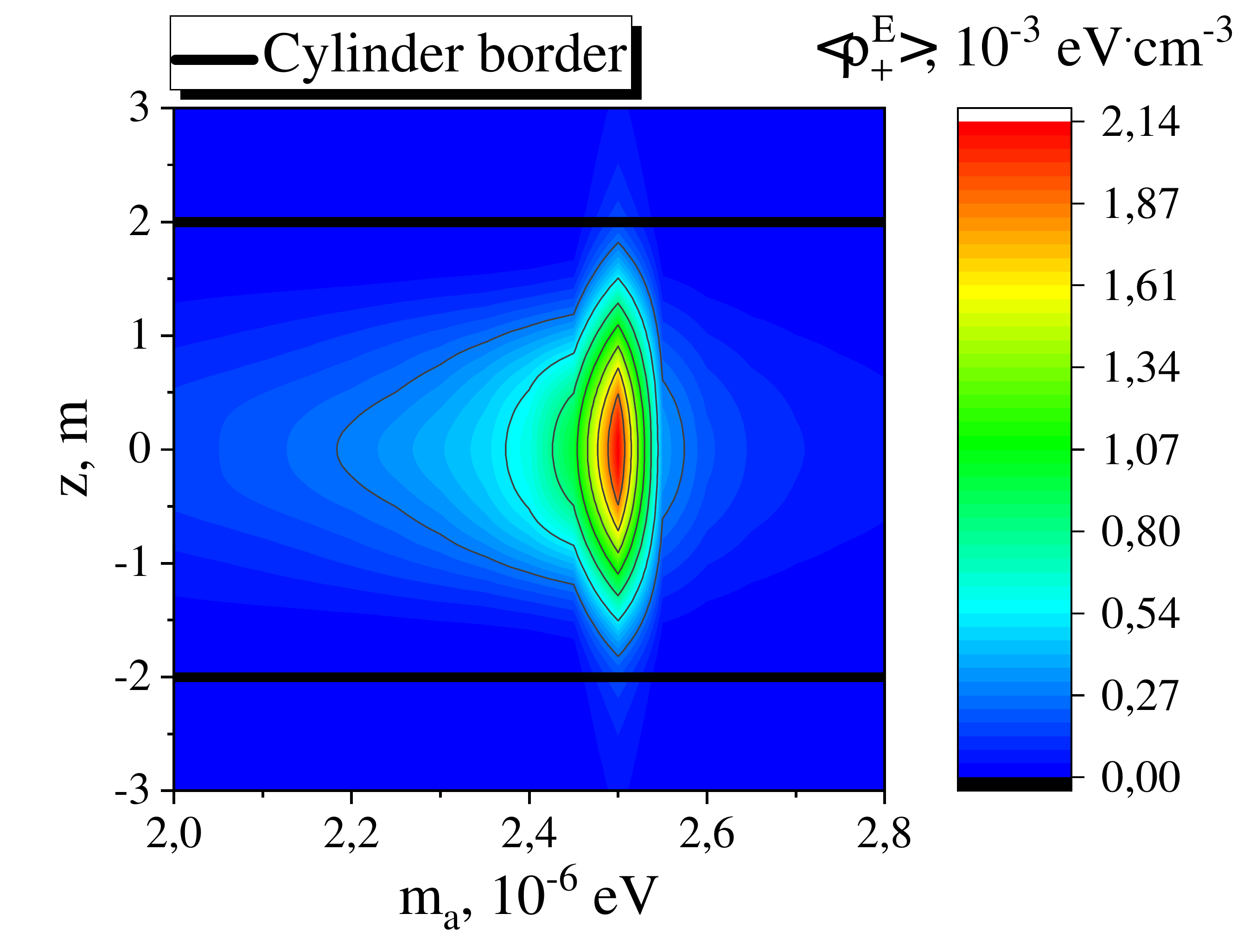}}
\end{minipage}
\hfill
\begin{minipage}[h]{0.49\linewidth}
\center{\includegraphics[width=1\linewidth]{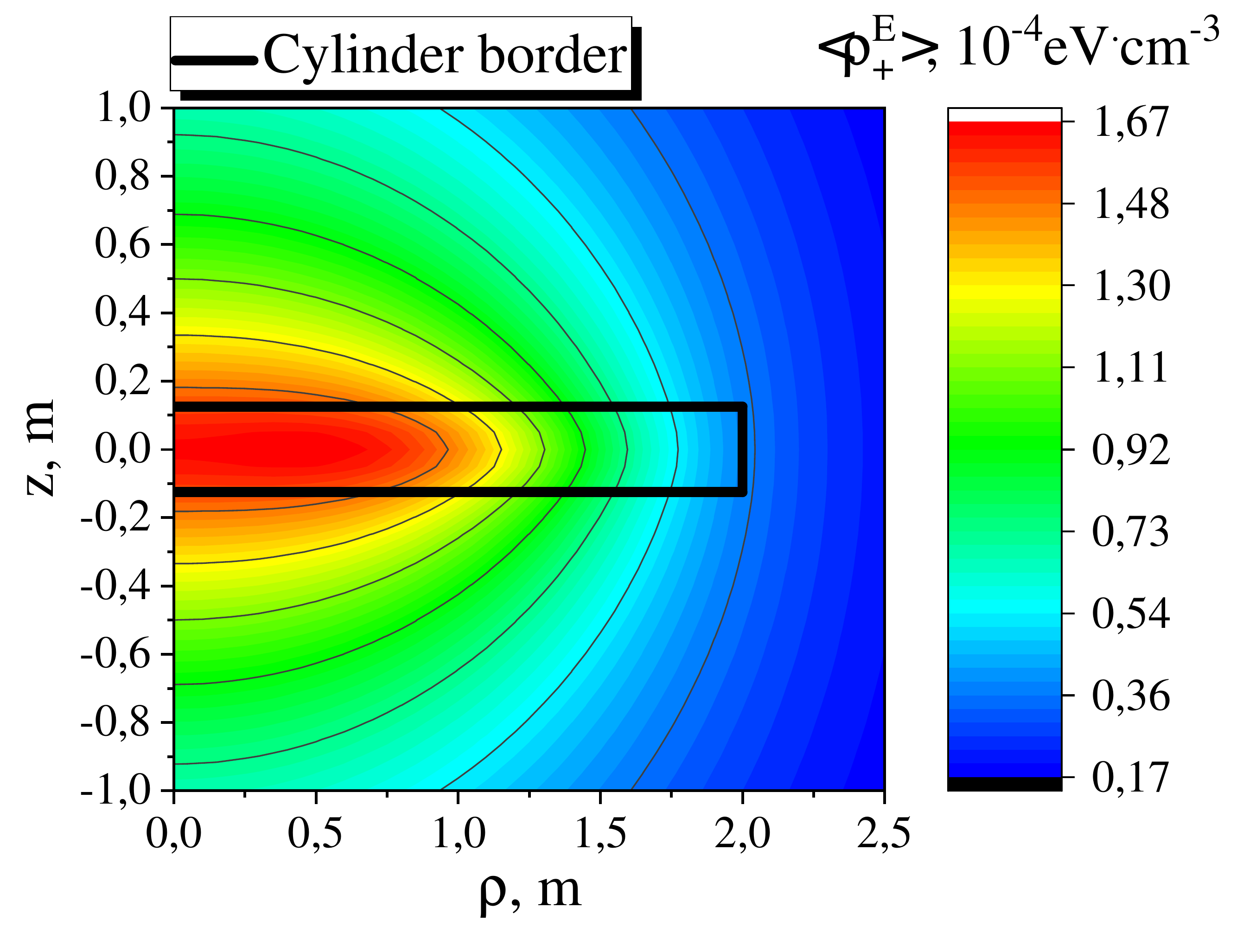}}
\center{\includegraphics[width=1\linewidth]{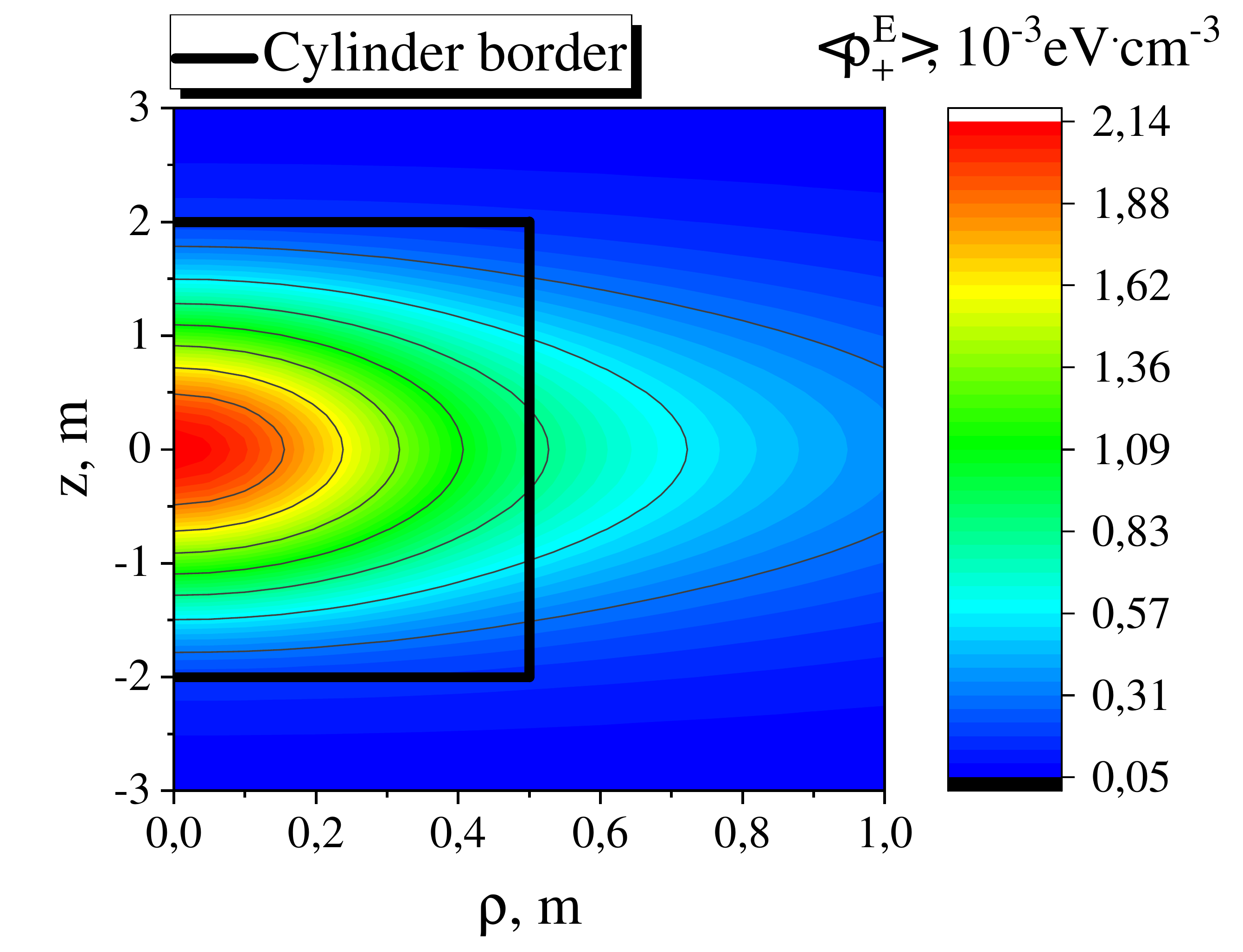}}
\end{minipage}
\caption{Contour plot for the time-averaged energy density $\braket{\rho^E_{+}}$ evaluated on the cylinder axis as function of axion mass $m_a$ and the distance $z$ from center of cavity (left panels) and spatial distribution of the time-averaged energy density $\braket{\rho^E_+}$ on the cavity section along its axis $(\rho,\,z)$ (right panels) with TM010+TE011 pump modes. Cavity dimensions: $L = 0.25$ m, $R = 2$ m, $\omega_+=27.8\cdot 10^{-7}$ eV (top), and $L = 4$ m, $R = 0.5$ m, $\omega_+=25.0\cdot 10^{-7}$ eV (bottom).}\label{fig:C2}
\end{figure}

\begin{table}[b]
\begin{center}
\begin{tabular}{cccccccccc} 
   \hline
   \hline
   & \multicolumn{5}{c}{$\text{Cavities parameters}$} & & \multirow{2}{*}{$\kappa_c$} &  \multirow{2}{*}{$g_{a\gamma\gamma}, \mathrm{GeV}^{-1}$ } & \\
   \cline{2-6}  
   & $R_1, \mathrm{m}$ & $L_1, \mathrm{m}$ & $\text{Prod. modes}$ & $\omega_+, \mathrm{eV}$ & $R_2, \mathrm{m}$ &  &  &  \\
   \hline
   \hline   
   & \multirow{10}{*}{$1$}  &  \multirow{10}{*}{$1$} & TM010+TE011 & $1.47 \cdot 10^{-6}$ & $0.33$ &  & $9.71 \cdot 10^{-3}$ & $1.55 \cdot 10^{-11}$  \\
   \cline{4-6} \cline{8-9}
   & & & TM011+TE012 & $2.26 \cdot 10^{-6}$ & $0.21$ &  & $4.72 \cdot 10^{-3}$ & $2.75\cdot 10^{-11}$  \\
   \cline{4-6} \cline{8-9}
   & & & TM012+TE013 & $3.38 \cdot 10^{-6}$ & $0.14$ &  & $4.07 \cdot 10^{-3}$ & $3.62\cdot 10^{-11}$  \\
   \cline{4-6} \cline{8-9}
   & & & TM020+TE011 & $2.10 \cdot 10^{-6}$ & $0.23$ &  & $5.45 \cdot 10^{-3}$ & $2.46\cdot 10^{-11}$  \\
   \cline{4-6} \cline{8-9}
   &  &  & TM011+TE011 & $1.78 \cdot 10^{-6}$ & $0.27$ & & $6.60 \cdot 10^{-4}$ & $6.53 \cdot 10^{-11}$  \\
    \cline{4-6} \cline{8-9}
   &  &  & TM021+TE011 & $2.26 \cdot 10^{-6}$ & $0.21$ & & $5.87 \cdot 10^{-4}$ & $7.79 \cdot 10^{-11}$  \\
    \cline{4-6} \cline{8-9}
   &  &  & TM021+TE021 & $2.81 \cdot 10^{-6}$ & $0.17 $ & & $7.42\cdot 10^{-4}$ & $7.72 \cdot 10^{-11}$  \\
    \cline{4-6} \cline{8-9}
   &  &  & TM013+TE014 & $4.57 \cdot 10^{-6}$ & $0.11$ & & $3.89\cdot 10^{-3}$ & $4.30 \cdot 10^{-11}$  \\
    \cline{4-6} \cline{8-9}
   &  &  & TM013+TE015 & $5.18 \cdot 10^{-6}$ & $0.09$ & & $7.88 \cdot 10^{-4}$ & $1.02 \cdot 10^{-10}$  \\
    \cline{4-6} \cline{8-9}
   &  &  & TM110+TE111 & $1.49 \cdot 10^{-6}$ & $0.32$ & & $2.85 \cdot 10^{-2} $ & $2.88 \cdot 10^{-11}$  \\
   \hline
   \hline
    & \multirow{10}{*}{$2$}  &  \multirow{10}{*}{$0.25$} & TM010+TE011 & $2.78 \cdot 10^{-6}$ & $0.17$ &  & $3.61 \cdot 10^{-3}$ & $3.49 \cdot 10^{-11}$  \\
   \cline{4-6} \cline{8-9}
    &  &  & TM011+TE012 & $7.57 \cdot 10^{-6}$ & $0.06$ & & $3.01 \cdot 10^{-3}$ & $6.29 \cdot 10^{-11}$  \\
    \cline{4-6} \cline{8-9}
    &  &  & TM012+TE013 & $1.26 \cdot 10^{-5}$ & $0.04$ & & $3.00 \cdot 10^{-3}$ & $8.14 \cdot 10^{-11}$  \\
    \cline{4-6} \cline{8-9}
     &  &  & TM020+TE011 & $3.09 \cdot 10^{-6}$ & $0.16$ & & $2.16 \cdot 10^{-3}$ & $4.76 \cdot 10^{-11}$  \\
    \cline{4-6} \cline{8-9}
   &  &  & TM011+TE011 & $5.07 \cdot 10^{-6}$ & $0.09$ & & $2.60 \cdot 10^{-6}$ & $1.75 \cdot 10^{-9}$  \\
    \cline{4-6} \cline{8-9}
   &  &  & TM021+TE011 & $5.12 \cdot 10^{-6}$ & $0.09$ & & $4.67 \cdot 10^{-6}$ & $1.31 \cdot 10^{-9}$  \\
    \cline{4-6} \cline{8-9}
   &  &  & TM021+TE021 & $5.18 \cdot 10^{-6}$ & $0.09$ & & $7.20\cdot 10^{-6}$ & $1.07 \cdot 10^{-9}$  \\
    \cline{4-6} \cline{8-9}
   &  &  & TM013+TE014 & $1.76 \cdot 10^{-5}$ & $0.03$ & & $2.99\cdot 10^{-3}$ & $9.63 \cdot 10^{-11}$  \\
    \cline{4-6} \cline{8-9}
   &  &  & TM013+TE015 & $2.01 \cdot 10^{-5}$ & $0.02$ & & $7.01 \cdot 10^{-5}$ & $6.72 \cdot 10^{-10}$  \\
    \cline{4-6} \cline{8-9}
   &  &  & TM110+TE111 & $2.90 \cdot 10^{-6}$ & $0.17$ & & $1.65 \cdot 10^{-3} $ & $5.26 \cdot 10^{-11}$  \\
   \hline
   \hline
   & \multirow{10}{*}{$0.5$}  &  \multirow{10}{*}{$4$} & TM010+TE011 & $2.50 \cdot 10^{-6}$ & $0.19$ &  & $4.53\cdot 10^{-3}$ & $2.95 \cdot 10^{-11}$  \\
   \cline{4-6} \cline{8-9}
   &  &  & TM011+TE012 & $2.54 \cdot 10^{-6}$ & $0.19$ & & $3.66 \cdot 10^{-3}$ & $3.31 \cdot 10^{-11}$  \\
    \cline{4-6} \cline{8-9}
     &  &  & TM012+TE013 & $2.62 \cdot 10^{-6}$ & $0.18$ & & $3.35 \cdot 10^{-3}$ & $3.51 \cdot 10^{-11}$  \\
    \cline{4-6} \cline{8-9}
    &  &  & TM020+TE011 & $3.75 \cdot 10^{-6}$ & $0.13$ & & $2.52 \cdot 10^{-3}$ & $4.84 \cdot 10^{-11}$  \\
    \cline{4-6} \cline{8-9}
   &  &  & TM011+TE011 & $2.52 \cdot 10^{-6}$ & $0.19$ & & $1.18 \cdot 10^{-3}$ & $5.79 \cdot 10^{-11}$  \\
    \cline{4-6} \cline{8-9}
   &  &  & TM021+TE011 & $3.75 \cdot 10^{-6}$ & $0.13$ & & $6.69\cdot 10^{-4}$ & $9.41 \cdot 10^{-11}$  \\
    \cline{4-6} \cline{8-9}
   &  &  & TM021+TE021 & $5.02 \cdot 10^{-6}$ & $0.10$ & & $7.47\cdot 10^{-4}$ & $1.03 \cdot 10^{-10}$  \\
    \cline{4-6} \cline{8-9}
   &  &  & TM013+TE014 & $2.73 \cdot 10^{-6}$ & $0.18$ & & $3.23\cdot 10^{-3}$ & $3.65 \cdot 10^{-11}$  \\
    \cline{4-6} \cline{8-9}
   &  &  & TM013+TE015 & $2.79 \cdot 10^{-6}$ & $0.17$ & & $1.91 \cdot 10^{-3}$ & $4.81 \cdot 10^{-11}$  \\
    \cline{4-6} \cline{8-9}
   &  &  & TM110+TE111 & $2.29 \cdot 10^{-6}$ & $0.21$ & & $1.47 \cdot 10^{-3} $ & $4.96 \cdot 10^{-11}$  \\
   \hline
   \hline
   
\end{tabular}
\end{center}
\caption{Resonant sensitivities for $g_{a \gamma\gamma}$ for different combinations of pump modes. $R_2$ is the radius of detecting cavity. $g_{a\gamma\gamma}$ is the detectable lower bound.}
    \label{tab:my_label}
\end{table}

\end{document}